\documentclass[journal,twoside]{IEEEtran}
\usepackage{cite}

\usepackage{graphicx}
\usepackage{xcolor}

\usepackage{amsmath}
\usepackage{mathrsfs}
\usepackage{amsfonts}
\usepackage{hyperref}
\hypersetup{
    colorlinks=true,
    linkcolor=blue,
    filecolor=magenta,      
    urlcolor=cyan
}

\usepackage[detect-all]{siunitx}
\usepackage{lmodern}
\usepackage[english]{babel}
\newtheorem{theorem}{Proposition}
\allowdisplaybreaks

\begin{document}
%
\title{New Results on Single User Massive MIMO}
%
%
%

\author{Kasturi Vasudevan$^1$, 
        Surendra Kota$^1$, 
        Lov Kumar$^1$ and~Himanshu Bhusan Mishra$^2$
\thanks{$^1$Department of Electrical Engineering Indian Institute of Technology Kanpur
208016 India e-mail: \{vasu, skota, lovkr20\}@iitk.ac.in, $^2$Department of Electronics
Engineering Indian Institute of Technology (Indian School of Mines) Dhanbad 826004
India email: himanshu@iitism.ac.in.}
}

%
%

\markboth{Book Chapter Intech Open}%
{Vasudevan \MakeLowercase{\textit{et al.}}: New Results on SU-MMIMO}
%



\maketitle

\begin{abstract}
Achieving high bit rates is the main goal of wireless technologies like 5G and
beyond. This translates to obtaining high spectral efficiencies using large number
of antennas at the transmitter and receiver (single user massive multiple input
multiple output or SU-MMIMO). It is possible to have a large number of antennas in the
mobile handset at mm-wave frequencies in the range $30 - 300$ GHz due to the
small antenna size.
In this work, we investigate the bit-error-rate (BER) performance of SU-MMIMO in two
scenarios (a) using serially concatenated turbo code (SCTC) in uncorrelated channel and (b)
parallel concatenated turbo code (PCTC) in correlated channel. Computer simulation results
indicate that the BER is quite insensitive to re-transmissions and wide variations in the
number of transmit and receive antennas. Moreover, we have obtained a BER of $10^{-5}$
at an average signal-to-interference plus noise ratio (SINR) per bit of just 1.25 dB with
512 transmit and receive antennas ($512\times 512$ SU-MMIMO system) with a spectral
efficiency of 256 bits/transmission or 256 bits/sec/Hz in an uncorrelated channel.
Similar BER results have been obtained for SU-MMIMO using PCTC in correlated channel.
A semi-analytic approach to estimating the BER of a turbo code has been derived.
\end{abstract}

%
\IEEEpeerreviewmaketitle

\section{Introduction}
%
%
%
%
\IEEEPARstart{A}{s} wireless technologies evolve beyond 5G \cite{9144301,9757375,9939166},
there is a growing need to
attain peak data rates of about gigabits per second per user, which is required for
high definition video, remote surgery, autonomous vehicles, gaming and so on, while
at the same time
consuming minimum transmit power. This can only be achieved by using multiple antennas
at the transmitter and receiver \cite{9864042,10002368,10006402,10045774,10049425},
small constellations like quadrature shift keying (QPSK)
and powerful error correcting codes like turbo or low density parity check (LDPC) codes.
Having a large number of antennas in the mobile handset is feasible in mm-wave frequencies
\cite{8901159,9310258,9705623,9946839}
($30 - 300$ GHz) due to the small antenna size. The main concern about mm wave
communications has been its rather high attenuation in outdoor environments with
rain and snow
\cite{8761087}. Therefore, at least in the initial stages, mm wave could be deployed
indoors. The second issue relates to the poor penetration characteristics of mm wave
through walls, doors, windows and other materials. This points towards to usage of mm wave
\cite{8901159} in a single room, say a big auditorium or underground parking and so on.
Reconfigurable intelligent surface (RIS)
\cite{9087848,9424177,9721205,10.1007/978-981-16-7423-5_102} could be used to boost
the propagation of mm waves, both indoors and outdoors.

Most of the massive MIMO systems discussed in the literature are multi-user (MU)
\cite{6798744,6777306,6808541,6971234,6987288,7160668,7342925,7439790,MMIMO_Khwandah_2021},
that is, the base station has a large number of antennas and the mobile handset has only
a single antenna ($N_t=1$). A large number of users are served simultaneously by the
base station. A comparison between MU-MMIMO and SU-MMIMO is given in
Table~\ref{Tbl:MU_SU_MMIMO_Comp} \cite{KV_MMWS2021,Vasu_MMIMO_INGR_2022}.
\begin{table}[tbhp]
\centering
\caption{Comparison of MU-MMIMO and SU-MMIMO.}
\input{mmimo_comp1.pstex_t}
\label{Tbl:MU_SU_MMIMO_Comp}
\end{table}
The base station in MU-MMIMO uses beamforming to improve the signal-to-noise
ratio at the mobile handset. On the other hand, SU-MMIMO uses spatial
multiplexing to improve the spectral efficiency in the downlink and uplink.
The comparison between beamforming and spatial multiplexing is given in
Table~\ref{Tbl:Beam_vs_SM} \cite{KV_MMWS2021,Vasu_MMIMO_INGR_2022}.
\begin{table}[tbhp]
\centering
\caption{Comparison of beamforming and spatial multiplexing.}
\input{beam_vs_sm.pstex_t}
\label{Tbl:Beam_vs_SM}
\end{table}
The total transmit power of SU-MMIMO using uncoded QPSK versus MU-MMIMO using
$M$-ary QAM is shown in Table~\ref{Tbl:MMIMO_QPSK_Mary}. The minimum Euclidean
distance between symbols of all constellations is taken to be 2.
\begin{table*}[tbhp]
\centering
\caption{SU-MMIMO using QPSK vs MU-MMIMO using $M$-ary.}
\input{mmimo_qpsk_mary.pstex_t}
\label{Tbl:MMIMO_QPSK_Mary}
\end{table*}
The peak-to-average power ratio (PAPR) for SU-MMIMO using QPSK is compared with
MU-MMIMO using $M$-ary QAM in Table~\ref{Tbl:MMIMO_PAPR1} \cite{KV_MMWS2021}.
Of course in the case of frequency selective fading channels, OFDM needs to be used,
which would result in PAPR greater than 0 dB even for QPSK signalling.
\begin{table}[tbhp]
\centering
\caption{PAPR of SU-MMIMO using QPSK vs MU-MMIMO using $M$-ary.}
\input{mmimo_papr1.pstex_t}
\label{Tbl:MMIMO_PAPR1}
\end{table}
It is clear from Tables~\ref{Tbl:MU_SU_MMIMO_Comp} -- \ref{Tbl:MMIMO_PAPR1} that
technologies that use SU-MMIMO have a lot to gain. Moreover, since all transmit
antennas use the same carrier frequency, there is no increase in bandwidth.

SU-MMIMO with equal number of transmit and receive antennas is given in
\cite{KV_OpSigPJ2019,73ddc0ea-7d42-4fdd-969d-da08c8e4d0c0}. The probability of
erasure in MIMO-OFDM is presented in \cite{KV_SSID2020}. A practical SU-MMIMO
receiver with estimated channel, carrier frequency offset and timing is described
in \cite{Vasu_intech:2019,d4bbbdf0-7468-4727-9ebe-76d5e6160b64}. SU-MMIMO with
unequal number of transmit and receive antennas and precoding is discussed in
\cite{KV_ARCI2021,da1844bd-7ee2-4d99-b53b-2339010e03b0} and the case without
precoding in \cite{KV_Oct_2021,4466bded-2b5a-454b-8f3d-a6dd0b74831d}. All the earlier
research on SU-MMIMO involved the use of a parallel concatenated turbo code (PCTC) and
uncorrelated channel. In this work, we investigate the performance of SU-MMIMO using
(a) serial concatenated turbo code (SCTC) in uncorrelated channel and (b) PCTC in
correlated channel. Throughout this article we assume that the channel is known
perfectly at the receiver. Perfect carrier and timing synchronization is also
assumed.

This work is organized as follows. Section~\ref{Sec:SU_MMIMO_SCTC} discusses
SU-MMIMO with SCTC in uncorrelated channel, the procedure for bit-error-rate
(BER) estimation and computer simulation results.
Section~\ref{Sec:SU_MMIMO_PCTC_Corr_Chan} deals with SU-MMIMO using PCTC in
correlated channel with and without precoding along with computer simulation
results. Section~\ref{Sec:Conclude} presents the conclusions and scope for
future work.
\section{SU-MMIMO with SCTC}
\label{Sec:SU_MMIMO_SCTC}
\subsection{System Model}
\label{SSec:SU_MMIMO_SCTC_Model}
\begin{figure*}[tbhp]
\centering
\input{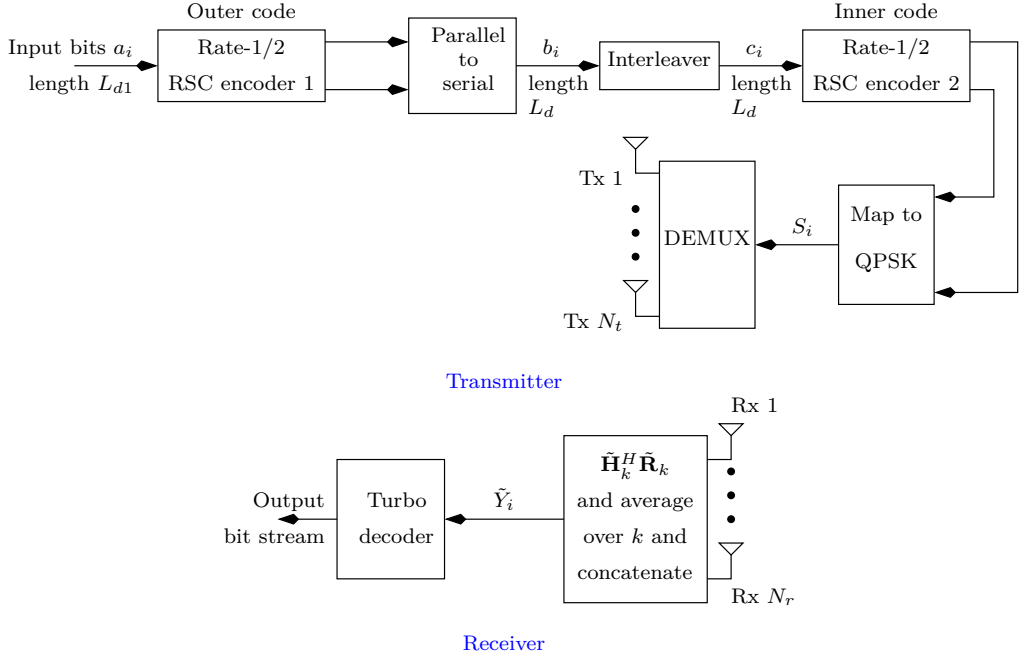}
\caption{SU-MMIMO with serially concatenated turbo code.}
\label{Fig:System_SCTC}
\end{figure*}

Consider the block diagram in Figure~\ref{Fig:System_SCTC}
\cite{Vasu07,KV_Oct_2021}. The input bits $a_i$, $1\le i \le L_{d1}$ is passed
through an outer rate-$1/2$ recursive systematic convolutional (RSC) encoder to
obtain the coded bit stream $b_i$, $1\le i \le L_d$, where
\begin{equation}
\label{Eq:SU_MMIMO_SCTC_Eq1}
L_d = 2 L_{d1}.
\end{equation}
Now $b_i$ is input to an interleaver to generate $c_i$, $1\le i \le L_d$. Next
$c_i$ is passed through an inner rate-$1/2$ RSC encoder and mapped to symbols
$S_i$, $1\le i \le L_d$, in a quadrature phase shift keyed (QPSK) constellation
having symbol coordinates $\pm 1\pm\,\mathrm{j}$, where $\mathrm{j}=\sqrt{-1}$.
Throughout this article we assume that bit ``0'' maps to $+1$ and bit ``1''
maps to $-1$.
The set of $L_d$ QPSK symbols constitute a ``frame'' and are transmitted using
$N_t$ antennas. We assume that
\begin{equation}
\label{Eq:SU_MMIMO_SCTC_Eq2}
\frac{L_d}{N_t} = \mbox{an integer}
\end{equation}
so that all symbols in the frame are transmitted using $N_t$ antennas. The set
of QPSK symbols transmitted simultaneously using $N_t$ antennas constitute a
``block''. The generator matrix for both the inner and outer rate-$1/2$ RSC
encoder is given by
\begin{equation}
\label{Eq:SU_MMIMO_SCTC_Eq2_1}
\mathbf{G}(D) = \left[
                \begin{array}{cc}
                 1 & \frac{1+D^2}{1+D+D^2}
                \end{array}
                \right].
\end{equation}
Hence, both encoders have $S_E=4$ states in the trellis.
Assuming uncorrelated Rayleigh flat fading, the received signal for the $k^{th}$
re-transmission ($0\le k\le N_{rt}-1$, $k$ is an integer) is given by (2) of
\cite{KV_Oct_2021}, which is repeated here for convenience
\begin{equation}
\label{Eq:SU_MMIMO_SCTC_Eq3}
\tilde{\mathbf{R}}_k = \tilde{\mathbf{H}}_k
                       \mathbf{S} +
                       \tilde{\mathbf{W}}_k
\end{equation}
where $\mathbf{S}\in \mathbb{C}^{N_t\times 1}$ whose elements are drawn
from the QPSK constellation, $\tilde{\mathbf{H}}_k\in \mathbb{C}^{N_r\times N_t}$
whose elements are mutually independent and $\mathscr{CN}(0,\, 2\sigma^2_H)$ and
and $\tilde{\mathbf{W}}_k\in \mathbb{C}^{N_r\times 1}$ is the additive 
white Gaussian noise (AWGN) vector whose elements are mutually independent
and $\mathscr{CN}(0,\, 2\sigma^2_W)$. Note that $\sigma^2_H,\, \sigma^2_W$
denote the variance per dimension (real part or imaginary part) and $N_r$ is
the number of receive antennas. We assume that $\tilde{\mathbf{H}}_k$ and
$\tilde{\mathbf{W}}_k$ are independent across blocks and re-transmissions,
hence (4) in  \cite{KV_OpSigPJ2019} is valid with $N$ replaced by $N_t$.
Recall that (see also (16) of \cite{KV_Oct_2021})
\begin{equation}
\label{Eq:SU_MMIMO_SCTC_Eq3_1}
N_{\mathrm{tot}} = N_t + N_r.
\end{equation}
Following the procedure
given in Section~4 of \cite{KV_Oct_2021} we get (see (36) of \cite{KV_Oct_2021})
\begin{equation}
\label{Eq:SU_MMIMO_SCTC_Eq4}
\tilde{Y}_i  =  F_i S_i + \tilde{U}_i \qquad \mbox{for $1\le i \le N_t$}.
\end{equation}
After concatenation over blocks, $\tilde{Y}_i$ in (\ref{Eq:SU_MMIMO_SCTC_Eq4})
for $1\le i \le L_d$ is sent to the turbo decoder (see also the sentence after
(25) in \cite{KV_OpSigPJ2019}). For the sake of consistency with earlier work
\cite{Vasu07}, we re-index $i$ as $0\le i \le L_d-1$ and use the
same index $i$ for $a_i$, $b_i$, $c_i$ and $Y_i$ without any ambiguity.
In the next subsection, we
discuss the turbo decoding (BCJR) algorithm \cite{BCJR74,Vasu_Book10} for the
inner code.
\subsection{BCJR for the Inner Code}
\label{SSec:BCJR_Inner_Code}
Let $\mathscr{D}_n$ denote the set of states that diverge from state $n$ in the
trellis \cite{Vasu07,Vasu_Book10}. Similarly, let $\mathscr{C}_n$ denote the set
of states that converge to state $n$. Let $\alpha_{i,\, n}$ denote the forward
sum-of-products (SOP) at time $i$, $0 \le i \le L_d - 2$, at state $n$,
$0 \le n \le S_E - 1$. Then the forward SOP can be recursively computed
as follows (see also (30) of \cite{Vasu07}):
\begin{align}
\label{Eq:SU_MMIMO_SCTC_Eq5}
\alpha_{i+1,\, n}' & = \sum_{m\in \mathscr{C}_n}
                       \alpha_{i,\, m}
                       \gamma_{i,\, m,\, n}
                        P(c_{i,\, m,\, n})    \nonumber  \\
\alpha_{0,\, n}    & =  1                     \nonumber  \\
\alpha_{i+1,\, n}  & = \left.
                       \alpha_{i+1,\, n}'
                       \middle/
                       \left(
                       \sum_{n=0}^{S_E-1}
                       \alpha_{i+1,\, n}'
                       \right)
                       \right.
\end{align}
where $P(c_{i, m, n})$ denotes the \textit{a priori} probability of the
systematic bit corresponding to the transition from encoder state $m$
to $n$, at time $i$ (this is set to 0.5 at the beginning of the first iteration).
The last equation in (\ref{Eq:SU_MMIMO_SCTC_Eq5}) is required to prevent
numerical instabilities \cite{Vasu_Book10}. We have
\begin{equation}
\label{Eq:SU_MMIMO_SCTC_Eq6}
\gamma_{i,\, m,\, n}   = \exp
                         \left(
                         -
                         \frac{\left(\tilde{Y}_i-S_{m,\, n}\right)^2}
                              {2\sigma^2_U}
                         \right)
\end{equation}
where $\tilde{Y}_i$ is given by (\ref{Eq:SU_MMIMO_SCTC_Eq4}), $S_{m,\, n}$ is the
QPSK symbol corresponding to the transition from encoder state $m$ to $n$ and
$\sigma^2_U$ is given by (38) of \cite{KV_Oct_2021} which is repeated here for
convenience:
\begin{align}
\label{Eq:SU_MMIMO_SCTC_Eq7}
 E
\left[
\left|
\tilde{U}_i
\right|^2
\right] & = \frac{8 \sigma_H^4 N_r (N_t-1) + 4\sigma_W^2\sigma^2_H N_r}
                 {N_{rt}}         \nonumber  \\
        &   \stackrel{\Delta}{=}
            \sigma^2_U.
\end{align}
Robust turbo decoding (see section 4.2 of \cite{Vasudevan2015}) can be employed
to compute $\gamma_{i,\, m,\, n}$ in (\ref{Eq:SU_MMIMO_SCTC_Eq6}).
Similarly, let $\beta_{i,\, m}$ denote the backward SOP at time $i$,
$1\le i \le L_d -1$, at state $m$, $0\le m \le S_E -1$. Then the
backward SOP can be recursively computed as (see also (33) of \cite{Vasu07}):
\begin{align}
\label{Eq:SU_MMIMO_SCTC_Eq8}
\beta_{i,\, m}'  & = \sum_{n\in \mathscr{D}_m}
                     \beta_{i+1,\, n}
                     \gamma_{i,\, m,\, n}
                      P(c_{i,\, m,\, n})    \nonumber  \\
\beta_{L_d,\, m} & =  1                     \nonumber  \\
\beta_{i,\, m}   & = \left.
                     \beta_{i,\, m}'
                     \middle/
                     \left(
                     \sum_{m=0}^{S_E-1}
                     \beta_{i,\, m}'
                     \right)
                     \right.
\end{align}
Let $\rho^+(n)$ denote the state that is reached from encoder state $n$
when the input symbol is $+1$. Similarly let $\rho^-(n)$ denote the
state that can be reached from encoder state $n$ when the input symbol
is $-1$. Then for $0\le i \le L_d-1$ we compute
\begin{align}
\label{Eq:SU_MMIMO_SCTC_Eq9}
C_{i+}                  & = \sum_{n=0}^{S_E-1}
                            \alpha_{i,\, n}
                            \gamma_{i,\, n,\,\rho^+(n)}
                            \beta_{i+1,\,\rho^+(n)}          \nonumber  \\
C_{i-}                  & = \sum_{n=0}^{S_E-1}
                            \alpha_{i,\, n}
                            \gamma_{i,\, n,\,\rho^-(n)}
                            \beta_{i+1,\,\rho^-(n)}.
\end{align}
Finally, the extrinsic information that is fed to the BCJR algorithm for the
outer code is computed as, for $0\le i \le L_d-1$, (see (36) of \cite{Vasu07}):
\begin{align}
\label{Eq:SU_MMIMO_SCTC_Eq10}
 E
\left(
 c_i = +1
\right) & =  C_{i+}/(C_{i+}+C_{i-})                \nonumber  \\
  E
\left(
 c_i = -1
\right) & =  C_{i-}/(C_{i+}+C_{i-}).
\end{align}
Next, we describe the BCJR for the outer code.
\subsection{BCJR for the Outer Code}
\label{SSec:BCJR_Outer_Code}
Let $\alpha_{i,\, n}$ denote the forward SOP at time $i$, $0\le i \le L_{d1}-2$,
at state $n$, $0\le n \le S_E -1$. Then the forward SOP is recursively computed
as follows:
\begin{align}
\label{Eq:SU_MMIMO_SCTC_Eq11}
\alpha_{i+1,\, n}' & = \sum_{m\in \mathscr{C}_n}
                       \alpha_{i,\, m}
                       \gamma_{\mathrm{sys},\, i,\, m,\, n}
                       \gamma_{\mathrm{par},\, i,\, m,\, n}
                        P(a_{i,\, m,\, n})    \nonumber  \\
\alpha_{0,\, n}    & =  1                     \nonumber  \\
\alpha_{i+1,\, n}  & = \left.
                       \alpha_{i+1,\, n}'
                       \middle/
                       \left(
                       \sum_{n=0}^{S_E-1}
                       \alpha_{i+1,\, n}'
                       \right)
                       \right.
\end{align}
where $P(a_{i,\, m,\, n})$ denotes the \textit{a priori} probability of the
systematic bit corresponding to the transition from state $m$ to
state $n$, at time $i$. In the absence of any other information, we assume
$P(a_{i,\, m,\, n})=0.5$ \cite{Singer02}. We also have for $0\le i \le L_{d1}-1$
(similar to (38) of \cite{Vasu07})
\begin{align}
\label{Eq:SU_MMIMO_SCTC_Eq12}
\gamma_{\mathrm{sys},\, i,\, m,\, n}
& = \left
    \{
    \begin{array}{ll}
     E\left(c_{\pi(2i)}=+1\right) &  \mbox{if $\mathscr{H}_1$}\\
     E\left(c_{\pi(2i)}=-1\right) &  \mbox{if $\mathscr{H}_2$}
    \end{array}
    \right.                                     \nonumber  \\
\gamma_{\mathrm{par},\, i,\, m,\, n}
& = \left
    \{
    \begin{array}{ll}
     E\left(c_{\pi(2i+1)}=+1\right) &  \mbox{if $\mathscr{H}_3$}\\
     E\left(c_{\pi(2i+1)}=-1\right) &  \mbox{if $\mathscr{H}_4$}
    \end{array}
    \right.
\end{align}
where $\pi(\cdot)$ denotes the interleaver map and
\begin{align}
\label{Eq:SU_MMIMO_SCTC_Eq13}
\mathscr{H}_1 & : \mbox{systematic bit from state $m$ to $n$ is $+1$} \nonumber  \\
\mathscr{H}_2 & : \mbox{systematic bit from state $m$ to $n$ is $-1$} \nonumber  \\
\mathscr{H}_3 & : \mbox{parity bit from state $m$ to $n$ is $+1$}     \nonumber  \\
\mathscr{H}_4 & : \mbox{parity bit from state $m$ to $n$ is $-1$}.
\end{align}
Observe that in (\ref{Eq:SU_MMIMO_SCTC_Eq12}) and (\ref{Eq:SU_MMIMO_SCTC_Eq13})
it is assumed that after the parallel-to-serial conversion in
Figure~\ref{Fig:System_SCTC}, $b_{2i}$ corresponds to the systematic (data) bits
and $b_{2i+1}$ corresponds to the parity bits for $0\le i\le L_{d1}-1$.

Similarly, let $\beta_{i,\, m}$ denote the backward SOP at time $i$,
$1\le i \le L_{d1}-1$, at state $m$, $0\le m \le S_E-1$. Then the backward
SOP can be recursively computed as:
\begin{align}
\label{Eq:SU_MMIMO_SCTC_Eq14}
\beta_{i,\, m}'  & = \sum_{n\in \mathscr{D}_m}
                     \beta_{i+1,\, n}
                     \gamma_{\mathrm{sys},\, i,\, m,\, n}
                     \gamma_{\mathrm{par},\, i,\, m,\, n}
                      P(a_{i,\, m,\, n})    \nonumber  \\
\beta_{L_{d1},\, m}
                 & =  1                     \nonumber  \\
\beta_{i,\, m}   & = \left.
                     \beta_{i,\, m}'
                     \middle/
                     \left(
                     \sum_{m=0}^{S_E-1}
                     \beta_{i,\, m}'
                     \right).
                     \right.
\end{align}
Next, for $0\le i\le L_{d1}-1$ we compute
\begin{align}
\label{Eq:SU_MMIMO_SCTC_Eq15}
B_{2i+}                 & = \sum_{n=0}^{S_E-1}
                            \alpha_{i,\, n}
                            \gamma_{\mathrm{par},\, i,\, n,\,\rho^+(n)}
                            \beta_{i+1,\,\rho^+(n)}          \nonumber  \\
B_{2i-}                 & = \sum_{n=0}^{S_E-1}
                            \alpha_{i,\, n}
                            \gamma_{\mathrm{par},\, i,\, n,\,\rho^-(n)}
                            \beta_{i+1,\,\rho^-(n)}.
\end{align}
Let $\mu^+(n)$ and $\mu^-(n)$ denote the states that are
reached from state $n$ when the parity bit is $+1$ and $-1$, respectively.
Similarly for $0\le i \le L_{d1}-1$ compute
\begin{align}
\label{Eq:SU_MMIMO_SCTC_Eq16}
B_{2i+1+}               & = \sum_{n=0}^{S_E-1}
                            \alpha_{i,\, n}
                            \gamma_{\mathrm{sys},\, i,\, n,\,\mu^+(n)}
                            \beta_{i+1,\,\mu^+(n)}          \nonumber  \\
B_{2i+1-}               & = \sum_{n=0}^{S_E-1}
                            \alpha_{i,\, n}
                            \gamma_{\mathrm{sys},\, i,\, n,\,\mu^-(n)}
                            \beta_{i+1,\,\mu^-(n)}.
\end{align}
The extrinsic information that is sent to the inner decoder for
$0\le i\le L_d-1$ is computed as
\begin{align}
\label{Eq:SU_MMIMO_SCTC_Eq17}
 E
\left(
 b_i = +1
\right) & =  B_{i+}/(B_{i+}+B_{i-})                \nonumber  \\
  E
\left(
 b_i = -1
\right) & =  B_{i-}/(B_{i+}+B_{i-})
\end{align}
where $B_{i+},\, B_{i-}$ are given by (\ref{Eq:SU_MMIMO_SCTC_Eq15}) or
(\ref{Eq:SU_MMIMO_SCTC_Eq16}) depending on whether $i$ is even or odd
respectively. Note that $P(c_{i,\, m,\, n})$ for $0\le i \le L_d-1$ in
(\ref{Eq:SU_MMIMO_SCTC_Eq5}) and (\ref{Eq:SU_MMIMO_SCTC_Eq8}) is equal to
\begin{equation}
\label{Eq:SU_MMIMO_SCTC_Eq18}
 P
\left(
 c_{i,\, m,\, n}
\right) = \left
          \{
          \begin{array}{ll}
           E
          \left(
           b_{\pi^{-1}(i)} = +1
          \right) &  \mbox{if $\mathscr{H}_1$}\\
           E
          \left(
           b_{\pi^{-1}(i)} = -1
          \right) &  \mbox{if $\mathscr{H}_2$}
          \end{array}
          \right.
\end{equation}
where $\pi^{-1}(\cdot)$ denotes the inverse interleaver map.
Note that $c_{i,\, m,\, n}$ are the systematic (data) bits for the inner
encoder.

After the convergence of the BCJR algorithm in the last iteration, the final
\textit{a posteriori} probabilities of $a_i$ for $0\le i\le L_{d1}-1$ is given
by
\begin{align}
\label{Eq:SU_MMIMO_SCTC_Eq19}
 P
\left(
 a_i = +1
\right) & =  E
            \left(
             b_{2i} = +1
            \right)
             E
            \left(
             c_{\pi(2i)} = +1
            \right)                       \nonumber  \\
 P
\left(
 a_i = -1
\right) & =  E
            \left(
             b_{2i} = -1
            \right)
             E
            \left(
             c_{\pi(2i)} = -1
            \right)
\end{align}
where $E\left(c_i=\pm 1\right)$ and $E\left(b_i=\pm 1\right)$ are given by
(\ref{Eq:SU_MMIMO_SCTC_Eq10}) and (\ref{Eq:SU_MMIMO_SCTC_Eq17}) respectively.
Finally note that for $0\le i \le L_{d1} - 1$
\begin{align}
\label{Eq:SU_MMIMO_SCTC_Eq19_1}
a_i = b_{2i} = c_{\pi(2i)}.
\end{align}
In the next section we present the estimation of the bit-error-rate (BER) of the
SCTC.
\subsection{Estimation of BER}
\label{SSec:Est_BER}
\begin{figure*}[tbhp]
\centering
\input{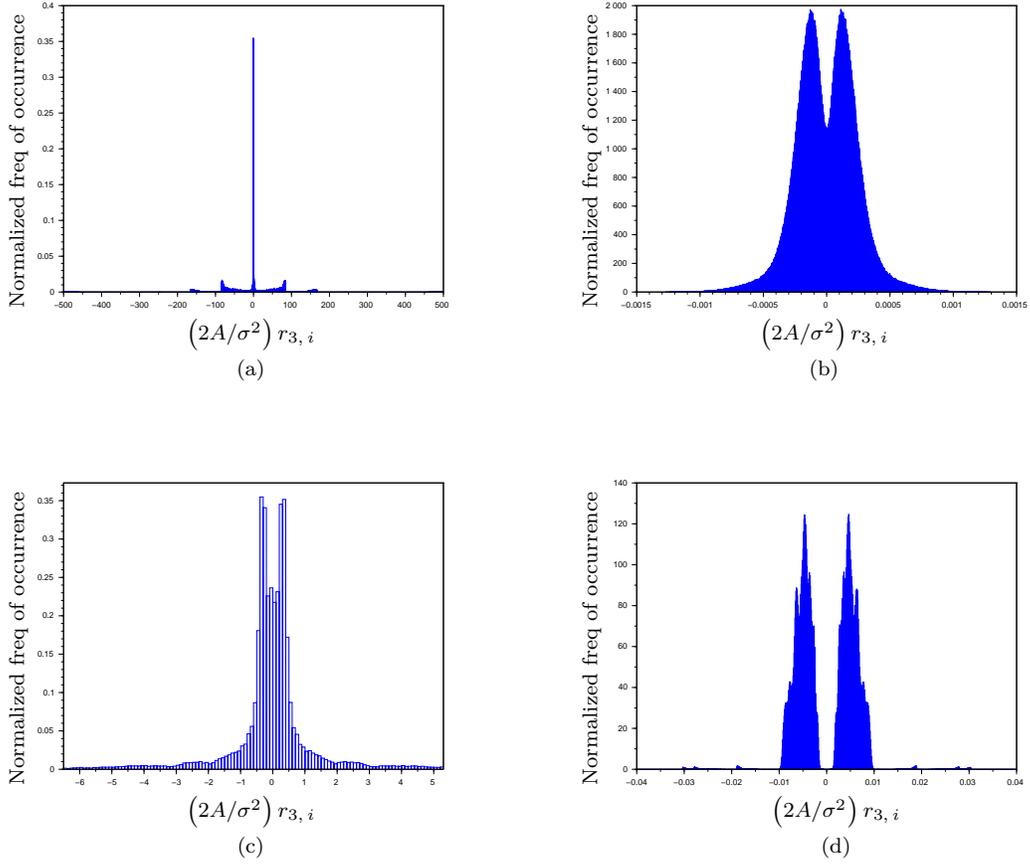}
\caption{Normalized histogram for $N_{\mathrm{tot}}=1024$, $N_t=512$, $N_{rt}=2$
         (a) $L_{d1}=1024$, $\mathrm{SNR}_{\mathrm{av},\, b}=1.25$ dB, $F=10^5$
             frames
         (b) $L_{d1}=50176$, $\mathrm{SNR}_{\mathrm{av},\, b}=0.3$ dB, $F=2000$
             frames
         (c) Expanded view of (a) around $r_{3,\, i}=0$ and
         (d) $L_{d1}=50176$, $\mathrm{SNR}_{\mathrm{av},\, b}=0.5$ dB, $F=2000$
             frames.}
\label{Fig:DMT55_i_Avg_Hist}
\end{figure*}
\begin{figure*}[tbhp]
\centering
\input{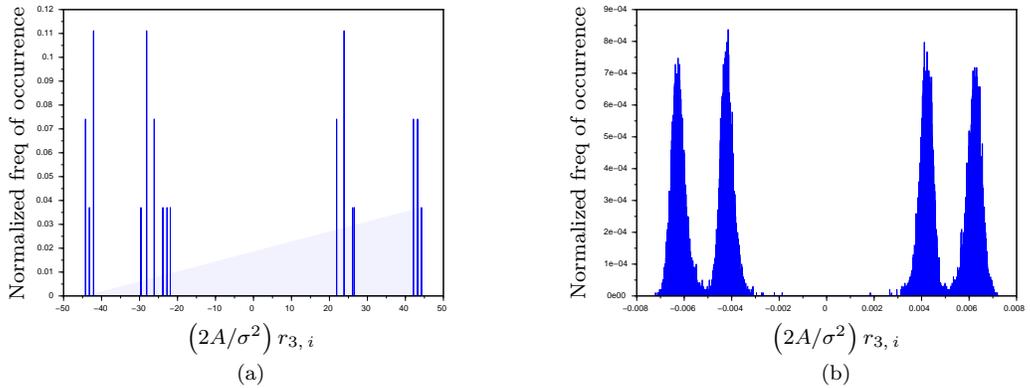}
\caption{Normalized histogram over two frames ($F=2$) for
         $N_{\mathrm{tot}}=1024$, $N_t=512$, $N_{rt}=2$
         (a) $L_{d1}=1024$, $\mathrm{SNR}_{\mathrm{av},\, b}=1.25$ dB and
         (b) $L_{d1}=50176$, $\mathrm{SNR}_{\mathrm{av},\, b}=0.5$ dB.}
\label{Fig:DMT55_i_Hist_Max_Fr2}
\end{figure*}
The estimation of BER of SCTC is based on the following propositions:
\begin{theorem}
\textit{The extrinsic information as computed in (\ref{Eq:SU_MMIMO_SCTC_Eq10}) and
(\ref{Eq:SU_MMIMO_SCTC_Eq17}) lies in the range $[0,\, 1]$ (0 and 1 included).
The extrinsic information in the range $(0,\, 1)$, 0 and 1 excluded,
is Gaussian distributed \cite{tenBrink01} for each frame.}
\end{theorem}
This is illustrated in Figure~\ref{Fig:DMT55_i_Avg_Hist} for different values
of the frame length $L_{d1}$, over many frames ($F$). We find that for large
values of $L_{d1}$, the histogram better approximates the Gaussian characteristic. 
It may be noted that the extrinsic information at the output of one decoder
is equal to the \textit{a priori} probabilities for the other decoder.
\begin{theorem}
\textit{After convergence of the BCJR algorithm in the final iteration, the
extrinsic information at a decoder output has the same mean and
variance as that of the \textit{a priori} probability at its input.}
\end{theorem}
\begin{theorem}
\textit{The mean and variance of the Gaussian distribution may vary from
frame to frame.}
\end{theorem}
This is illustrated in Figure~\ref{Fig:DMT55_i_Hist_Max_Fr2} over two frames,
that is, $F=2$.

Based on \textit{Propositions 1 \& 2} and (\ref{Eq:SU_MMIMO_SCTC_Eq19_1}),
after convergence of the BCJR algorithm,
we can write for $0\le i \le L_{d1} - 1$
\begin{align}
\label{Eq:SU_MMIMO_SCTC_Eq20}
 E
\left(
 b_{2i}=+1
\right) & = \frac{1}{\sigma\sqrt{2\pi}}
            \mathrm{e}^{-(r_{1,\, i}-A)^2/(2\sigma^2)}     \nonumber  \\
 E
\left(
 c_{\pi(2i)}=+1
\right) & = \frac{1}{\sigma\sqrt{2\pi}}
            \mathrm{e}^{-(r_{2,\, i}-A)^2/(2\sigma^2)}
\end{align}
where it is assumed that bit ``0'' maps to $A$ and bit ``1'' maps to $-A$ and
\begin{align}
\label{Eq:SU_MMIMO_SCTC_Eq21}
r_{1,\, i} & = \pm A + w_{1,\, i}                     \nonumber  \\
r_{2,\, i} & = \pm A + w_{2,\, i}
\end{align}
where $w_{1,\, i},\, w_{2,\, i}$ denote real-valued samples of zero-mean
additive white Gaussian noise (AWGN) with variance $\sigma^2$. Similarly
we have
\begin{align}
\label{Eq:SU_MMIMO_SCTC_Eq22}
 E
\left(
 b_{2i}=-1
\right) & = \frac{1}{\sigma\sqrt{2\pi}}
            \mathrm{e}^{-(r_{1,\, i}+A)^2/(2\sigma^2)}     \nonumber  \\
 E
\left(
 c_{\pi(2i)}=-1
\right) & = \frac{1}{\sigma\sqrt{2\pi}}
            \mathrm{e}^{-(r_{2,\, i}+A)^2/(2\sigma^2)}.
\end{align}
Clearly
\begin{align}
\label{Eq:SU_MMIMO_SCTC_Eq23}
\ln
\left(
\frac{
 E
\left(
 b_{2i}=+1
\right)}{E\left(b_{2i}=-1\right)}
\right)
        & = \frac{2A}{\sigma^2} r_{1,\, i}             \nonumber  \\
\ln
\left(
\frac{
 E
\left(
 c_{\pi(2i)}=+1
\right)}{E\left(c_{\pi(2i)}=-1\right)}
\right)
        & = \frac{2A}{\sigma^2} r_{2,\, i}.
\end{align}
From (\ref{Eq:SU_MMIMO_SCTC_Eq19}) and (\ref{Eq:SU_MMIMO_SCTC_Eq23}) we have
for $0\le i \le L_{d1}-1$
\begin{align}
\label{Eq:SU_MMIMO_SCTC_Eq24}
\ln
\left(
\frac{
 P
\left(
 a_i=+1
\right)}{P\left(a_i=-1\right)}
\right) & = \frac{2A}{\sigma^2}
            \left(
             r_{1,\, i} + r_{2,\, i}
            \right)                    \nonumber  \\
        &   \stackrel{\Delta}{=}
            \frac{2A}{\sigma^2}
             r_{3,\, i}.
\end{align}
Consider the average
\begin{align}
\label{Eq:SU_MMIMO_SCTC_Eq25}
\mathscr{Y}
  & = \frac{2A}{\sigma^2L_{d2}}
      \sum_{i=0}^{L_{d2}-1}
       a_i
       r_{3,\, i}                              \nonumber  \\
  & = \frac{4A^2}{\sigma^2} + \mathscr{Z}
\end{align}
where
\begin{align}
\label{Eq:SU_MMIMO_SCTC_Eq26}
\mathscr{Z}
& =   \frac{2A}{\sigma^2L_{d2}}
      \sum_{i=0}^{L_{d2}-1}
       a_i
      \left(
       w_{1,\, i} + w_{2,\, i}
      \right)                   \nonumber  \\
L_{d2}
& \le  L_{d1}.
\end{align}
Note that the average in
(\ref{Eq:SU_MMIMO_SCTC_Eq25}) is done over less than $L_{d1}$ terms to avoid
situations like
\begin{equation}
\label{Eq:SU_MMIMO_SCTC_Eq26_1}
 P
\left(
 a_i=\pm 1
\right) = \mbox{1 or 0}.
\end{equation}
In fact, only those time instants $i$ have been considered in the
summation of (\ref{Eq:SU_MMIMO_SCTC_Eq25}) for which
\begin{equation}
\label{Eq:SU_MMIMO_SCTC_Eq26_2}
 P
\left(
 a_i=\pm 1
\right)  > \mathrm{e}^{-500}.
\end{equation}
Now
\begin{align}
\label{Eq:SU_MMIMO_SCTC_Eq27}
 E
\left[
\mathscr{Z}
\right]
   & =  0                               \nonumber  \\
 E
\left[
\mathscr{Z}^2
\right]
   & = \frac{4A^2}{\sigma^4L_{d2}^2}
       \sum_{i=0}^{L_{d2}-1}
        2
       \sigma^2                         \nonumber  \\
   & = \frac{8A^2}{\sigma^2L_{d2}}
\end{align}
where we have used the fact that $w_{1,\, i},\, w_{2,\, i}$ are independent.
Now, we know that the probability of error for the BPSK signal in
(\ref{Eq:SU_MMIMO_SCTC_Eq24}), that is
\begin{equation}
\label{Eq:SU_MMIMO_SCTC_Eq28}
r_{3,\, i} = r_{1,\, i}+r_{2,\, i} = \pm 2A + w_{1,\, i} + w_{2,\, i}
\end{equation}
is equal to \cite{Vasu_Book10}
\begin{equation}
\label{Eq:SU_MMIMO_SCTC_Eq29}
P(e) = \frac{1}{2}
       \mbox{erfc}
       \left(
       \sqrt{
       \frac{A^2}{\sigma^2}}
       \right).
\end{equation}
Therefore from (\ref{Eq:SU_MMIMO_SCTC_Eq25}), (\ref{Eq:SU_MMIMO_SCTC_Eq27}) and
(\ref{Eq:SU_MMIMO_SCTC_Eq29}) we have
\begin{equation}
\label{Eq:SU_MMIMO_SCTC_Eq30}
P_f(e)
       \approx
       \frac{1}{2}
       \mbox{erfc}
       \left(
       \sqrt{
       \frac{\left|\mathscr{Y}\right|}{4}}
       \right)
\end{equation}
where $P_f(e)$ denotes the probability of bit error for frame ``$f$'' and
\begin{equation}
\label{Eq:SU_MMIMO_SCTC_Eq31}
 E
\left[
\mathscr{Z}^2
\right] \rightarrow 0 \qquad \mbox{for $L_{d2}\gg 1$.}
\end{equation}
Observe that it is necessary to take the absolute value of $\mathscr{Y}$ in
(\ref{Eq:SU_MMIMO_SCTC_Eq30}) since there is a possibility that it can be
negative.
The average probability of bit error over $F$ frames is given by
\begin{equation}
\label{Eq:SU_MMIMO_SCTC_Eq32}
P(e) = \frac{1}{F}
       \sum_{f=0}^{F-1}
        P_f(e).
\end{equation}
In the next section we present computer simulation results
for SU-MMIMO using SCTC in uncorrelated channel.
\subsection{Simulation Results}
\label{SSec:SCTC_Uncorr_Results}
\begin{table}[tbhp]
\centering
\caption{Simulation parameters for results in
         Figures~\ref{Fig:Ber_NUM_ANT1024_Uncorr_DMT55i} --
                 \ref{Fig:Ber_NUM_ANT2_Uncorr_DMT55i}}
\input{sim_param_sctc_dmt55i.pstex_t}
\label{Tbl:Sim_Param_SCTC_DMT55i}
\end{table}
\begin{figure*}[tbhp]
\centering
\input{ber_num_ant1024_uncorr_dmt55i.pstex_t}
\caption{Simulation results for $N_{\mathrm{tot}}=1024$.}
\label{Fig:Ber_NUM_ANT1024_Uncorr_DMT55i}
\end{figure*}
\begin{figure*}[tbhp]
\centering
\input{ber_num_ant32_uncorr_dmt55i.pstex_t}
\caption{Simulation results for $N_{\mathrm{tot}}=32$.}
\label{Fig:Ber_NUM_ANT32_Uncorr_DMT55i}
\end{figure*}
\begin{figure}[tbhp]
\centering
\input{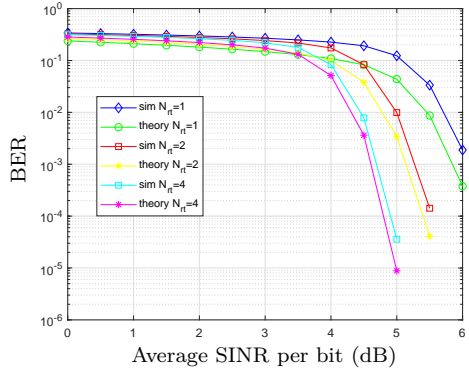}
\caption{Simulation results for $N_{\mathrm{tot}}=2$, $N_t=1$.}
\label{Fig:Ber_NUM_ANT2_Uncorr_DMT55i}
\end{figure}
The simulation parameters are given in Table~\ref{Tbl:Sim_Param_SCTC_DMT55i}.
We can make the following observations from
Figures~\ref{Fig:Ber_NUM_ANT1024_Uncorr_DMT55i} --
        \ref{Fig:Ber_NUM_ANT2_Uncorr_DMT55i} \cite{KV_Oct_2021}:
\begin{itemize}
    \item The theoretical prediction of BER closely matches with simulations.
    \item For  $N_{\mathrm{tot}}=32,\, 1024$, the BER is quite insensitive to
          wide variations in the total number
          of antennas $N_{\mathrm{tot}}$, transmit antennas $N_t$ and
          retransmissions $N_{rt}$.
    \item For $N_{\mathrm{tot}}=2$, the BER improves significantly with
          increasing retransmissions.
\end{itemize}
In Figure~\ref{Fig:Ber_NUM_ANT1024_Uncorr_DMT55i}(c) we observe that there
is more than 1 dB improvement in SINR compared to
Figures~\ref{Fig:Ber_NUM_ANT1024_Uncorr_DMT55i}(a, b),
\ref{Fig:Ber_NUM_ANT32_Uncorr_DMT55i} and
\ref{Fig:Ber_NUM_ANT2_Uncorr_DMT55i}. However, large values of $L_{d1}$ may
introduce more latency which is contrary to the requirements of 5G and beyond.
In the next section we present SU-MMIMO using PCTC in correlated channel.
\section{SU-MMIMO using PCTC in Correlated Channel}
\label{Sec:SU_MMIMO_PCTC_Corr_Chan}
\subsection{System Model}
\label{SSec:SU_MMIMO_PCTC_Corr_Chan_Sys_Model}
The block diagram of the system is identical to Figure~2 in \cite{KV_Oct_2021}
and the received signal is given by (\ref{Eq:SU_MMIMO_SCTC_Eq3}).
Note that in (\ref{Eq:SU_MMIMO_SCTC_Eq3}), the channel autocorrelation matrix is given by
\begin{align}
\label{Eq:EE677A_Asgn1_Eq1}
 \mathbf{R}_{\tilde{\mathbf{H}}\tilde{\mathbf{H}}}
& = \frac{1}{2}
     E
    \left[
    \tilde{\mathbf{H}}_k^H\tilde{\mathbf{H}}_k
    \right]                                                   \nonumber  \\
& =  N_r
    \mathbf{I}_{N_t}
\end{align}
where the superscript ``$H$'' denotes Hermitian and $\mathbf{I}_{N_t}$ denotes the
$N_t\times N_t$ identity matrix. In this section, we investigate the situation where
$\mathbf{R}_{\tilde{\mathbf{H}}\tilde{\mathbf{H}}}$ is not an identity matrix, but is
a valid autocorrelation matrix \cite{Vasu_Book10}.
As mentioned in \cite{KV_Oct_2021}, the elements of $\tilde{\mathbf{H}}_k$ -- given by
$\tilde{H}_{k,\, i,\, j}$ for the $k^{th}$ re-transmission, $i^{th}$ row, $j^{th}$
column of $\tilde{\mathbf{H}}_k$ -- are zero-mean, complex Gaussian random variables with
variance per dimension equal to $\sigma^2_H$. The in-phase and quadrature components of
$\tilde{H}_{k,\, i,\, j}$ -- denoted by $H_{k,\, i,\, j,\, I}$ and
$H_{k,\, i,\, j,\, Q}$ respectively -- are statistically independent. Moreover,
we assume that the rows of $\tilde{\mathbf{H}}_k$ are statistically independent.
Following the procedure in \cite{KV_Oct_2021} for the case without precoding, we
now find the expression for the average SINR per bit before and after averaging over
re-transmissions ($k$). All symbols and notations have the usual meaning, as
given in \cite{KV_Oct_2021}.
\subsection{SINR Analysis}
\label{SSec:SU_MMIMO_PCTC_Corr_Chan_SINR_Ana}
The $i^{th}$ element of $\tilde{\mathbf{H}}_k^H\tilde{\mathbf{R}}_k$ is given by
(25) of \cite{KV_Oct_2021} which is repeated here for convenience
\begin{equation}
\label{Eq:EE677A_Asgn1_Sol_Eq1}
\tilde{Y}_{k,\, i} = \tilde{F}_{k,\, i,\, i} S_i +
                     \tilde{I}_{k,\, i} +
                     \tilde{V}_{k,\, i}
                     \quad \mbox{for $1\leq i\leq N_t$}
\end{equation}
where
\begin{align}
\label{Eq:EE677A_Asgn1_Sol_Eq2}
\tilde{V}_{k,\, i}      & = \sum_{j=1}^{N_r}
                            \tilde{H}_{k,\, j,\, i}^*
                            \tilde W_{k,\, j}                 \nonumber  \\ 
\tilde{I}_{k,\, i}      & = \sum_{\substack{j=1\\j\neq i}}^{N_t}
                            \tilde{F}_{k,\, i,\, j} S_j       \nonumber  \\
\tilde{F}_{k,\, i,\, j} & = \sum_{l=1}^{N_r}
                            \tilde{H}_{k,\, l,\, i}^*
                            \tilde{H}_{k,\, l,\, j}.
\end{align}
We have
\begin{align}
\label{Eq:EE677A_Asgn1_Sol_Eq3}
 E
\left[
\tilde{F}_{k,\, i,\, i}^2
\right] & =  E
            \left[
            \sum_{l=1}^{N_r}
            \left|
            \tilde{H}_{k,\, l,\, i}
            \right|^2
            \sum_{m=1}^{N_r}
            \left|
            \tilde{H}_{k,\, m,\, i}
            \right|^2
            \right]                         \nonumber  \\
        & =  E
            \left[
            \sum_{l=1}^{N_r}
            \left(
             H_{k,\, l,\, i,\, I}^2 +
             H_{k,\, l,\, i,\, Q}^2
            \right)
            \right.                         \nonumber  \\
        &   \left.
            \qquad
            \sum_{m=1}^{N_r}
            \left(
             H_{k,\, m,\, i,\, I}^2 +
             H_{k,\, m,\, i,\, Q}^2
            \right)
            \right]                         \nonumber  \\
        & =  4\sigma_H^4 N_r(N_r+1)
\end{align}
which is identical to (27) in \cite{KV_Oct_2021} and we have used the following
properties         
\begin{enumerate}
    \item The in-phase and quadrature components of $\tilde{H}_{k,\, i,\, j}$ are
          independent.
    \item The rows of $\tilde{\mathbf{H}}_k$ are independent.
    \item For zero-mean, real-valued Gaussian random variable $X$ with variance equal to
          $\sigma^2_X$, $E\left[X^4\right]=3\sigma_X^4$.
\end{enumerate}
The interference power is
\begin{align}
\label{Eq:EE677A_Asgn1_Sol_Eq4}
 E
\left[
\left|
\tilde{I}_{k,\, i}
\right|^2
\right] & =  E
            \left[
            \sum_{\substack{j=1\\j\neq i}}^{N_t}
            \tilde{F}_{k,\, i,\, j} S_j
            \sum_{\substack{l=1\\l\neq i}}^{N_t}
            \tilde{F}_{k,\, i,\, l}^* S_l^*
            \right]                                          \nonumber  \\
        & = \sum_{\substack{j=1\\j\neq i}}^{N_t}
            \sum_{\substack{l=1\\l\neq i}}^{N_t}
             E
            \left[
            \tilde{F}_{k,\, i,\, j}
            \tilde{F}_{k,\, i,\, l}^*
            \right]
             E
            \left[
             S_j S_l^*
            \right]                                          \nonumber  \\
        & =  P_{\mathrm{av}}
            \sum_{\substack{j=1\\j\neq i}}^{N_t}
             E
            \left[
            \left| 
            \tilde{F}_{k,\, i,\, j}
            \right|^2
            \right].
\end{align}
where we have used (9) in \cite{KV_Oct_2021}. Similarly the noise power is
\begin{align}
\label{Eq:EE677A_Asgn1_Sol_Eq5}
 E
\left[
\left|
\tilde{V}_{k,\, i}
\right|^2
\right] & =  E
            \left[
            \sum_{j=1}^{N_r}
            \tilde{H}_{k,\, j,\, i}^* \tilde{W}_{k,\, j}
            \sum_{m=1}^{N_r}
            \tilde{H}_{k,\, m,\, i} \tilde{W}_{k,\, m}^*
            \right]                                           \nonumber  \\
        & = \sum_{j=1}^{N_r}
            \sum_{m=1}^{N_r}
             E
            \left[ 
            \tilde{H}_{k,\, j,\, i}^*
            \tilde{H}_{k,\, m,\, i}
            \right]
             E
            \left[
            \tilde{W}_{k,\, m}^* \tilde{W}_{k,\, j}
            \right]                                           \nonumber  \\
            & = \sum_{j=1}^{N_r}
            \sum_{m=1}^{N_r}
             2
            \sigma^2_H
            \delta_K(j-m)
             2
            \sigma^2_W(j-m)                                   \nonumber  \\
        & =  4 N_r
            \sigma^2_H
            \sigma^2_W
\end{align}
which is identical to (29) in \cite{KV_Oct_2021} and we have used the following
properties:
\begin{enumerate}
    \item Rows of $\tilde{\mathbf{H}}_k$ are independent.
    \item Sifting property of the Kronecker delta function.
    \item Noise and channel coefficients are independent.
\end{enumerate}
Now in (\ref{Eq:EE677A_Asgn1_Sol_Eq4})
\begin{align}
\label{Eq:EE677A_Asgn1_Sol_Eq6}
 E
\left[
\left|
\tilde{F}_{k,\, i,\, j}
\right|^2
\right] & =  E
            \left[
            \sum_{l=1}^{N_r}
            \tilde{H}_{k,\, l,\, i}^*
            \tilde{H}_{k,\, l,\, j}
            \sum_{m=1}^{N_r}
            \tilde{H}_{k,\, m,\, i}
            \tilde{H}_{k,\, m,\, j}^*
            \right]                                            \nonumber  \\
        & = \sum_{l=1}^{N_r}
             E
            \Biggl[
            \tilde{H}_{k,\, l,\, i}^*
            \tilde{H}_{k,\, l,\, j}
            \Biggl(
            \tilde{H}_{k,\, l,\, i}
            \tilde{H}_{k,\, l,\, j}^*
            \Biggr.
            \Biggr.                                            \nonumber  \\
        &   \qquad
            \Biggl.
            \Biggl. +
            \sum_{\substack{m=1\\m\ne l}}^{N_r}
            \tilde{H}_{k,\, m,\, i}
            \tilde{H}_{k,\, m,\, j}^*
            \Biggr)
            \Biggr]                                            \nonumber  \\           
        & = \sum_{l=1}^{N_r}
             E
            \Biggl[
            \left|
            \tilde{H}_{k,\, l,\, i}
            \right|^2
            \left|
            \tilde{H}_{k,\, l,\, j}
            \right|^2
            \Biggr.                                            \nonumber  \\
        &   \qquad    +
            \Biggl.
            \Biggl(
            \sum_{\substack{m=1\\m\ne l}}^{N_r}
            \tilde{H}_{k,\, l,\, i}^*
            \tilde{H}_{k,\, l,\, j}
            \tilde{H}_{k,\, m,\, i}
            \tilde{H}_{k,\, m,\, j}^*
            \Biggr)
            \Biggr].
\end{align}
Now the first summation in (\ref{Eq:EE677A_Asgn1_Sol_Eq6}) is equal to
\begin{align}
\label{Eq:EE677A_Asgn1_Sol_Eq7}
E_1 & =  E
        \left[
        \left|
        \tilde{H}_{k,\, l,\, i}
        \right|^2
        \left|
        \tilde{H}_{k,\, l,\, j}
        \right|^2
        \right]                                              \nonumber  \\
    & =  E
        \left[
        \left(
         H_{k,\, l,\, i,\, I}^2 +
         H_{k,\, l,\, i,\, Q}^2
        \right)
        \left(
         H_{k,\, l,\, j,\, I}^2 +
         H_{k,\, l,\, j,\, Q}^2
        \right)
        \right]                                              \nonumber  \\
    & =  4
        \sigma^4_H + 4 R_{\tilde{H}\tilde{H},\, j-i}^2
\end{align}
where we have used the property that for real-valued, zero-mean Gaussian
random variables $X_i$, $1\le i \le 4$ \cite{Papoulis91,Vasu_AC_PS}
\begin{equation}
\label{Eq:EE677A_Asgn1_Sol_Eq8}
 E
\left[
 X_1 X_2 X_3 X_4
\right] = C_{12}C_{34} + C_{13}C_{24} + C_{14}C_{23}
\end{equation}
where
\begin{equation}
\label{Eq:EE677A_Asgn1_Sol_Eq9}
 C_{ij} =
 E
\left[
 X_i X_j
\right] \qquad \mbox{for $1\le i,\, j\le 4$}
\end{equation}
and
\begin{align}
\label{Eq:EE677A_Asgn1_Sol_Eq10}
 R_{\tilde{H}\tilde{H},\, j-i}
& =
 E
\left[
 H_{k,\, l,\, i,\, I}
 H_{k,\, l,\, j,\, I}
\right]                                           \nonumber  \\
& =
 E
\left[
 H_{k,\, l,\, i,\, Q}
 H_{k,\, l,\, j,\, Q}
\right]                                           \nonumber  \\
& =
\frac{1}{2}
 E
\left[
\tilde{H}_{k,\, l,\, i}^*
\tilde{H}_{k,\, l,\, j}
\right]                                           \nonumber  \\
& =
 R_{\tilde{H}\tilde{H},\, i-j}
\end{align}
is the real-valued autocorrelation of $\tilde{H}_{k,\, m,\, n}$ and we have
made the assumption that the in-phase and quadrature components of
$\tilde{H}_{k,\, m,\, n}$ are independent. The second
summation in (\ref{Eq:EE677A_Asgn1_Sol_Eq6}) can be written as
\begin{align}
\label{Eq:EE677A_Asgn1_Sol_Eq11}
E_2
& =
\sum_{\substack{m=1\\m\ne l}}^{N_r}
 E
\left[
\tilde{H}_{k,\, l,\, i}^*
\tilde{H}_{k,\, l,\, j}
\tilde{H}_{k,\, m,\, i}
\tilde{H}_{k,\, m,\, j}^*
\right]                                                    \nonumber  \\
& =
\sum_{\substack{m=1\\m\ne l}}^{N_r}
 E
\left[
\tilde{H}_{k,\, l,\, i}^*
\tilde{H}_{k,\, l,\, j}
\right]
 E
\left[
\tilde{H}_{k,\, m,\, i}
\tilde{H}_{k,\, m,\, j}^*
\right]                                                    \nonumber  \\
& =
\sum_{\substack{m=1\\m\ne l}}^{N_r}
 4R_{\tilde{H}\tilde{H},\, j-i}^2                          \nonumber  \\
& =
 4(N_r-1)R_{\tilde{H}\tilde{H},\, j-i}^2
\end{align}
where we have used the property that the rows of $\tilde{\mathbf{H}}_k$ are
independent. Therefore (\ref{Eq:EE677A_Asgn1_Sol_Eq6}) becomes
\begin{align}
\label{Eq:EE677A_Asgn1_Sol_Eq12}
 E
\left[
\left|
\tilde{F}_{k,\, i,\, j}
\right|^2
\right] & =  N_r (E_1 +E_2)                           \nonumber  \\
        & =  4N_r
            \left[
            \sigma_H^4 + R_{\tilde{H}\tilde{H},\, j-i}^2 +
            \left(
             N_r-1
            \right)
             R_{\tilde{H}\tilde{H},\, j-i}^2
            \right]                                   \nonumber  \\
        & =  4N_r
            \left[
            \sigma_H^4 + N_r R_{\tilde{H}\tilde{H},\, j-i}^2
            \right].
\end{align}
The total power of interference plus noise is
\begin{align}
\label{Eq:EE677A_Asgn1_Sol_Eq13}
 E
\left[
\left|
\tilde{I}_{k,\, i} + \tilde{V}_{k,\, i}
\right|^2
\right] & =  E
            \left[
            \left|
            \tilde{I}_{k,\, i}
            \right|^2
            \right] +
             E
            \left[
            \left|
            \tilde{V}_{k,\, i}
            \right|^2
            \right]                                    \nonumber  \\
        & =  4 P_{\mathrm{av}}
             N_r
            \sum_{\substack{j=1\\j\ne i}}^{N_t}
            \left[
            \sigma^4_H + N_r R_{\tilde{H}\tilde{H},\, j-i}^2
            \right]                                    \nonumber  \\
        &   \qquad +
             4 N_r\sigma^2_H\sigma^2_W
\end{align}
where we have made the assumption that noise and symbols are independent.
The average SINR per bit for the $i^{th}$ transmit antenna is similar to (31)
of \cite{KV_Oct_2021} which is repeated here for convenience
\begin{equation}
\label{Eq:EE677A_Asgn1_Sol_Eq14}
\mathrm{SINR}_{\mathrm{av},\, b,\, i}
          = \frac{
             E
            \left[
            \left|
            \tilde{F}_{k,\, i,\, i} S_i
            \right|^2
            \right]
            \times 2N_{rt}}
            {
             E
            \left[
            \left|
            \tilde{I}_{k,\, i} + \tilde{V}_{k,\, i}
            \right|^2
            \right]
            }                       \qquad  \mbox{for $1\le i \le N_t$}
\end{equation}
into which (\ref{Eq:EE677A_Asgn1_Sol_Eq3}) and (\ref{Eq:EE677A_Asgn1_Sol_Eq13})
have to be substituted. The upper bound on the average SINR per bit for
the $i^{th}$ transmit antenna is obtained by
setting $\sigma^2_W=0$ in (\ref{Eq:EE677A_Asgn1_Sol_Eq13}),
(\ref{Eq:EE677A_Asgn1_Sol_Eq14}) and is given by, for $1\le i \le N_t$
\begin{equation}
\label{Eq:SU_MMIMO_SCTC_Eq33}
\mathrm{SINR}_{\mathrm{av},\, b,\, \mathrm{UB},\, i}
          = \frac{
            \sigma_H^4
            \left(
             1 + N_r
            \right)
            \times 2N_{rt}}
            {
            \sum_{\substack{j=1\\j\ne i}}^{N_t}
            \left[
            \sigma^4_H + N_r R_{\tilde{H}\tilde{H},\, j-i}^2
            \right]
            }.
\end{equation}
Observe that in contrast to (31) and (32) in \cite{KV_Oct_2021}, the average
SINR per bit and its upper bound depend on the transmit antenna. Let us now
compute the average SINR per bit after averaging over retransmissions. The
received signal after averaging over retransmissions is given by
(\ref{Eq:SU_MMIMO_SCTC_Eq4}) with (see also (20) of \cite{KV_Oct_2021})
\begin{align}
\label{Eq:SU_MMIMO_SCTC_Eq34}
F_i         & = \frac{1}{N_{rt}}
                \sum_{k=0}^{N_{rt}-1}
                \tilde{F}_{k,\, i,\, i}                        \nonumber  \\
\tilde{U}_i & = \frac{1}{N_{rt}}
                \sum_{k=0}^{N_{rt}-1}
                \left(
                \tilde{I}_{k,\, i} +
                \tilde{V}_{k,\, i}
                \right)                                        \nonumber  \\
            & = \frac{1}{N_{rt}}
                \sum_{k=0}^{N_{rt}-1}
                \tilde{U}_{k,\, i}'   \qquad \mbox{(say)}
\end{align}
where $\tilde{F}_{k,\, i,\, i}$, $\tilde{I}_{k,\, i}$ and $\tilde{V}_{k,\, i}$
are given in (\ref{Eq:EE677A_Asgn1_Sol_Eq1}). The power of the signal component
of (\ref{Eq:SU_MMIMO_SCTC_Eq4}) is
\begin{align}
\label{Eq:SU_MMIMO_SCTC_Eq35}
 E
\left[
\left|
 S_i
\right|^2
 F_i^2
\right] & =  P_{\mathrm{av}}
             E
            \left[
             F_i^2
            \right]                                     \nonumber  \\
        & = \frac{P_{\mathrm{av}}}{N_{rt}^2}
             E
            \left[
            \sum_{k=0}^{N_{rt}-1}
            \tilde{F}_{k,\, i,\, i}
            \sum_{l=0}^{N_{rt}-1}
            \tilde{F}_{l,\, i,\, i}
            \right]                                     \nonumber  \\
        & = \frac{P_{\mathrm{av}}}{N_{rt}^2}
            \sum_{k=0}^{N_{rt}-1}
            \Biggl[
            \sum_{\substack{l=0\\l\ne k}}^{N_{rt}-1}
             E[\tilde{F}_{k,\, i,\, i}]
             E[\tilde{F}_{l,\, i,\, i}]
            \Biggr.                                     \nonumber  \\
        &   \qquad +
            \Biggl.
             E
            \left[
            \left|
            \tilde{F}_{k,\, i,\, i}
            \right|^2
            \right]
            \Biggr]
\end{align}
where we have used the fact that the channel is independent across
retransmissions, therefore
\begin{equation}
\label{Eq:SU_MMIMO_SCTC_Eq36}
E[\tilde{F}_{k,\, i,\, i} \tilde{F}_{l,\, i,\, i}] =
E[\tilde{F}_{k,\, i,\, i}]
E[\tilde{F}_{l,\, i,\, i}] \qquad \mbox{for $k\ne l$}.
\end{equation}
Now
\begin{align}
\label{Eq:SU_MMIMO_SCTC_Eq37}
E[\tilde{F}_{k,\, i,\, i}]
                   & =  E
                       \left[
                       \sum_{l=1}^{N_r}
                       \left|
                       \tilde{H}_{k,\, l,\, i}
                       \right|^2
                       \right]                          \nonumber  \\
                   & =  2 N_r\sigma_H^2.
\end{align}
Substituting (\ref{Eq:EE677A_Asgn1_Sol_Eq3}) and (\ref{Eq:SU_MMIMO_SCTC_Eq37})
in (\ref{Eq:SU_MMIMO_SCTC_Eq35}) we get
\begin{equation}
\label{Eq:SU_MMIMO_SCTC_Eq38}
 E
\left[
\left|
 S_i
\right|^2
 F_i^2
\right] = \frac{4N_r P_{\mathrm{av}} \sigma_H^4}{N_{rt}}
          \left(
           1 + N_r N_{rt}
          \right).
\end{equation}
The power of the interference component in (\ref{Eq:SU_MMIMO_SCTC_Eq4}) and
(\ref{Eq:SU_MMIMO_SCTC_Eq34}) is
\begin{align}
\label{Eq:SU_MMIMO_SCTC_Eq39}
 E
\left[
\left|
\tilde{U}_i
\right|^2
\right] & =
          \frac{1}{N_{rt}^2}
           E
          \left[
          \sum_{k=0}^{N_{rt}-1}
          \left(
          \tilde{I}_{k,\, i} +
          \tilde{V}_{k,\, i}
          \right)
          \sum_{l=0}^{N_{rt}-1}
          \left(
          \tilde{I}_{l,\, i}^* +
          \tilde{V}_{l,\, i}^*
          \right)
          \right]                                  \nonumber  \\
        & =
          \frac{1}{N_{rt}^2}
          \sum_{k=0}^{N_{rt}-1}
          \sum_{l=0}^{N_{rt}-1}
           E
          \left[
          \tilde{I}_{k,\, i}
          \tilde{I}_{l,\, i}^*
          \right] +
           E
          \left[
          \tilde{V}_{k,\, i}
          \tilde{V}_{l,\, i}^*
          \right]
\end{align}
where we have used the following properties from (\ref{Eq:EE677A_Asgn1_Sol_Eq2})
\begin{align}
\label{Eq:SU_MMIMO_SCTC_Eq40}
 E
\left[
\tilde{I}_{k,\, i}
\right] & =  E
            \left[
            \tilde{V}_{k,\, i}
            \right]                       \nonumber  \\
        & =  0                            \nonumber  \\
 E
\left[
\tilde{I}_{k,\, i}
\tilde{V}_{l,\, i}^*
\right] & =  E
            \left[
            \tilde{V}_{k,\, i}
            \tilde{I}_{l,\, i}^*
            \right]                       \nonumber  \\
        & =  0    \qquad   \mbox{for all $k,\, l$}
\end{align}
since $S_j$ and $\tilde{W}_{k,\, j}$ are mutually independent with zero-mean.
Now
\begin{align}
\label{Eq:SU_MMIMO_SCTC_Eq41}
 E
\left[
\tilde{I}_{k,\, i}
\tilde{I}_{l,\, i}^*
\right] & =  E
            \left[
            \sum_{\substack{j=1\\j\ne i}}^{N_t}
            \tilde{F}_{k,\, i,\, j} S_j
            \sum_{\substack{n=1\\n\ne i}}^{N_t}
            \tilde{F}_{l,\, i,\, n}^* S_n^*
            \right]                                 \nonumber  \\
        & = \sum_{\substack{j=1\\j\ne i}}^{N_t}
            \sum_{\substack{n=1\\n\ne i}}^{N_t}
             E
            \left[
            \tilde{F}_{k,\, i,\, j}
            \tilde{F}_{l,\, i,\, n}^*
            \right]
             E
            \left[
             S_j S_n^*
            \right]                                 \nonumber  \\
        & = \sum_{\substack{j=1\\j\ne i}}^{N_t}
            \sum_{\substack{n=1\\n\ne i}}^{N_t}
             E
            \left[
            \tilde{F}_{k,\, i,\, j}
            \tilde{F}_{l,\, i,\, n}^*
            \right]
             P_{\mathrm{av}}
            \delta_K(j-n)                           \nonumber  \\
        & =  P_{\mathrm{av}}
            \sum_{\substack{j=1\\j\ne i}}^{N_t}
             E
            \left[
            \tilde{F}_{k,\, i,\, j}
            \tilde{F}_{l,\, i,\, j}^*
            \right]
\end{align}
where we have used the property that the symbols are uncorrelated and
$\delta_K(\cdot)$ is the Kronecker delta function \cite{Vasu_Book10}. When
$k=l$, (\ref{Eq:SU_MMIMO_SCTC_Eq41}) is given by (\ref{Eq:EE677A_Asgn1_Sol_Eq4})
and (\ref{Eq:EE677A_Asgn1_Sol_Eq12}). When $k\ne l$,
(\ref{Eq:SU_MMIMO_SCTC_Eq41}) is given by
\begin{align}
\label{Eq:SU_MMIMO_SCTC_Eq42}
 E
\left[
\tilde{I}_{k,\, i}
\tilde{I}_{l,\, i}^*
\right] & =  P_{\mathrm{av}}
            \sum_{\substack{j=1\\j\ne i}}^{N_t}
             E
            \left[
            \tilde{F}_{k,\, i,\, j}
            \right]
             E
            \left[
            \tilde{F}_{l,\, i,\, j}^*
            \right]                                     \nonumber  \\
        & =  P_{\mathrm{av}}
            \sum_{\substack{j=1\\j\ne i}}^{N_t}
             4
             N_r^2
             R_{\tilde{H}\tilde{H},\, j-i}^2
\end{align}
where we have used (\ref{Eq:EE677A_Asgn1_Sol_Eq2}) and
(\ref{Eq:EE677A_Asgn1_Sol_Eq10}). Similarly, we have
\begin{equation}
\label{Eq:SU_MMIMO_SCTC_Eq43}
 E
\left[
\tilde{V}_{k,\, i}
\tilde{V}_{l,\, i}^*
\right] = 4 N_r \sigma_H^2 \sigma^2_W \delta_K(k-l)
\end{equation}
where we have used (\ref{Eq:EE677A_Asgn1_Sol_Eq5}). Substituting
(\ref{Eq:EE677A_Asgn1_Sol_Eq4}), (\ref{Eq:EE677A_Asgn1_Sol_Eq12}),
(\ref{Eq:SU_MMIMO_SCTC_Eq42}) and (\ref{Eq:SU_MMIMO_SCTC_Eq43}) in
(\ref{Eq:SU_MMIMO_SCTC_Eq39}) we get
\begin{align}
\label{Eq:SU_MMIMO_SCTC_Eq44}
 E
\left[
\left|
\tilde{U}_i
\right|^2
\right] & = \frac{1}{N_{rt}^2}
            \left[
             4 P_{\mathrm{av}} N_r N_{rt}
            \sum_{\substack{j=1\\j\ne i}}^{N_t}
            \left(
            \sigma_H^4 + N_r R_{\tilde{H}\tilde{H},\, j-i}^2
            \right)
            \right.                                       \nonumber  \\
        &   \qquad +
            \left.
             4 P_{\mathrm{av}} N_r^2 N_{rt} (N_{rt}-1)
            \sum_{\substack{j=1\\j\ne i}}^{N_t}
             R_{\tilde{H}\tilde{H},\, j-i}^2
            \right]                                       \nonumber  \\
        &   \qquad +
            \frac{4N_r}{N_{rt}}
            \sigma_H^2
            \sigma^2_W                                    \nonumber  \\
        & = \frac{1}{N_{rt}}
            \left[
             4 P_{\mathrm{av}} N_r
            \sum_{\substack{j=1\\j\ne i}}^{N_t}
            \left(
            \sigma_H^4 + N_r R_{\tilde{H}\tilde{H},\, j-i}^2
            \right)
            \right.                                       \nonumber  \\
        &   \qquad +
            \left.
             4 P_{\mathrm{av}} N_r^2 (N_{rt}-1)
            \sum_{\substack{j=1\\j\ne i}}^{N_t}
             R_{\tilde{H}\tilde{H},\, j-i}^2
            \right]                                       \nonumber  \\
        &   \qquad +
            \frac{4N_r}{N_{rt}}
            \sigma_H^2
            \sigma^2_W.
\end{align}
The average SINR per bit for the $i^{th}$ transmit antenna, after averaging
over retransmissions (also referred to as ``combining'' \cite{KV_Oct_2021})
is given by
\begin{equation}
\label{Eq:SU_MMIMO_SCTC_Eq45}
\mathrm{SINR}_{\mathrm{av},\, b,\, C,\, i} =
\frac{2P_{\mathrm{av}}E\left[F_i^2\right]}
     {E\left[\left|\tilde{U}_i\right|^2\right]}
\end{equation}
into which (\ref{Eq:SU_MMIMO_SCTC_Eq38}) and (\ref{Eq:SU_MMIMO_SCTC_Eq44})
have to be substituted.
The upper bound on the average SINR per bit after ``combining'' for
the $i^{th}$ transmit antenna is given by
\begin{equation}
\label{Eq:SU_MMIMO_SCTC_Eq45_1}
\mathrm{SINR}_{\mathrm{av},\, b,\, C,\, \mathrm{UB},\, i} =
\left.
\mathrm{SINR}_{\mathrm{av},\, b,\, C,\, i}
\right|_{\sigma_W^2=0}.
\end{equation}
\begin{figure*}[tbhp]
\centering
\input{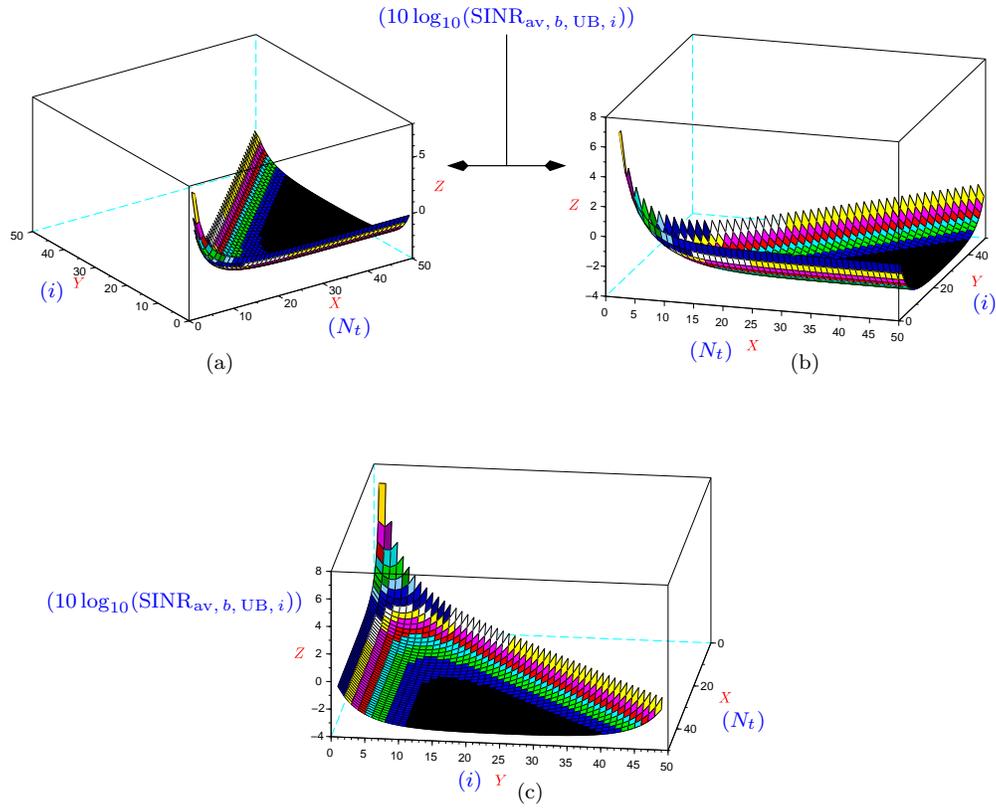}
\caption{Plot of $\mathrm{SINR}_{\mathrm{av},\, b,\,\mathrm{UB},\, i}$
         for $N_{\mathrm{tot}}=1024$, $N_{rt}=2$. (a) Back view. (b) Sideview.
         (c) Front view.}
\label{Fig:SINR_UB_BA_Numant1024}
\end{figure*}
\begin{figure*}[tbhp]
\centering
\input{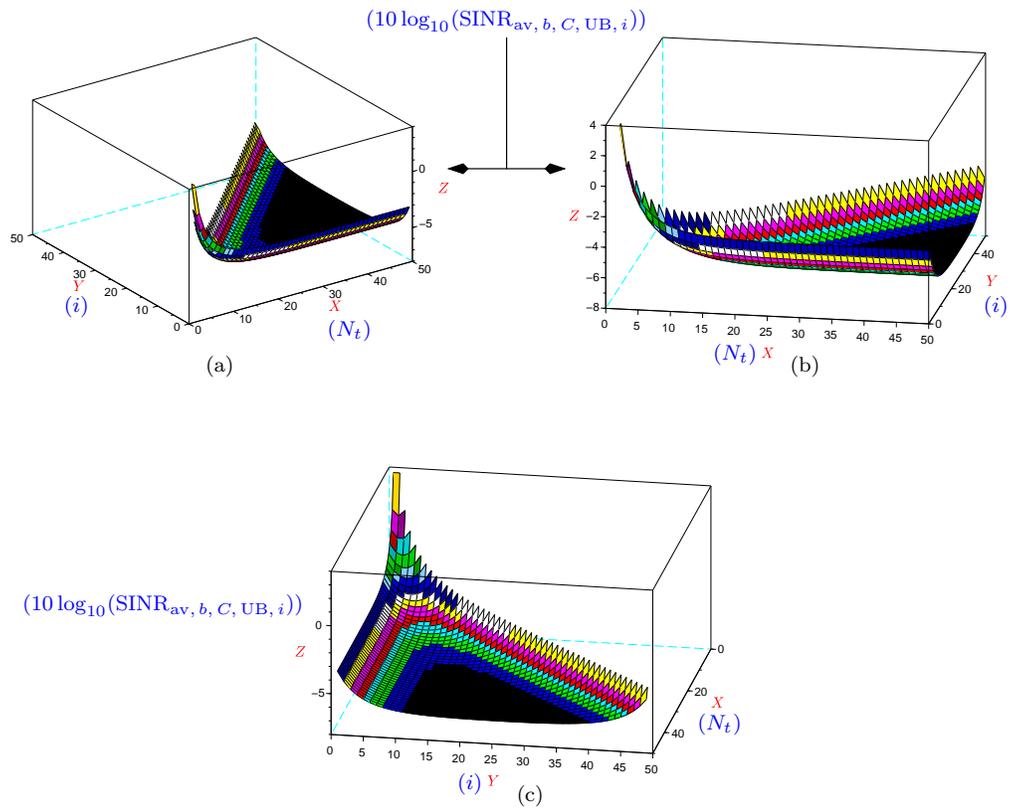}
\caption{Plot of $\mathrm{SINR}_{\mathrm{av},\, b,\, C,\,\mathrm{UB},\, i}$
         for $N_{\mathrm{tot}}=1024$, $N_{rt}=2$. (a) Back view. (b) Side view.
         (c) Front view.}
\label{Fig:SINR_UB_AA_Numant1024}
\end{figure*}
The plots of the average SINR per bit for the $i^{th}$ transmit antenna
before and after ``combining'' are shown in
Figures~\ref{Fig:SINR_UB_BA_Numant1024} and
\ref{Fig:SINR_UB_AA_Numant1024} respectively for $N_{\mathrm{tot}}=1024$ and
$N_{rt}=2$. The channel correlation is given by
\begin{equation}
\label{Eq:SU_MMIMO_SCTC_Eq46}
R_{\tilde{H}\tilde{H},\, j-i} = 0.9^{|j-i|} \sigma^2_H
\end{equation}
in (\ref{Eq:EE677A_Asgn1_Sol_Eq10}), which is obtained by passing samples of
white Gaussian noise through a unit-energy, first-order infinite impulse
response (IIR) lowpass filter with $a=-0.9$ (see (30) of \cite{Vasu_SPJ04}).

We observe in Figures~\ref{Fig:SINR_UB_BA_Numant1024} and
\ref{Fig:SINR_UB_AA_Numant1024} that
\begin{enumerate}
    \item The upper bound on the average SINR per bit decreases rapidly
          with increasing transmit
          antennas $N_t$ and falls below 0 dB for $N_t>5$
          (see Figures~\ref{Fig:SINR_UB_BA_Numant1024}(b) and
          \ref{Fig:SINR_UB_AA_Numant1024}(b)). Since the spectral
          efficiency of the system is $N_t/(2N_{rt})$ bits/sec/Hz (see (33) of
          \cite{KV_Oct_2021}), the system
          would be of no practical use, since the BER would be close to 0.5
          for $N_t>5$.
    \item The upper bound on the average SINR per bit after ``combining''
          is \textit{less} than
          that before ``combining''. Therefore retransmissions are
          ineffective.
\end{enumerate}
In view of the above observation, it becomes necessary to design a better
receiver using precoding. This is presented in the next section.
\subsection{Precoding}
\label{SSec:Precoding}
\begin{figure*}[tbhp]
\centering
\input{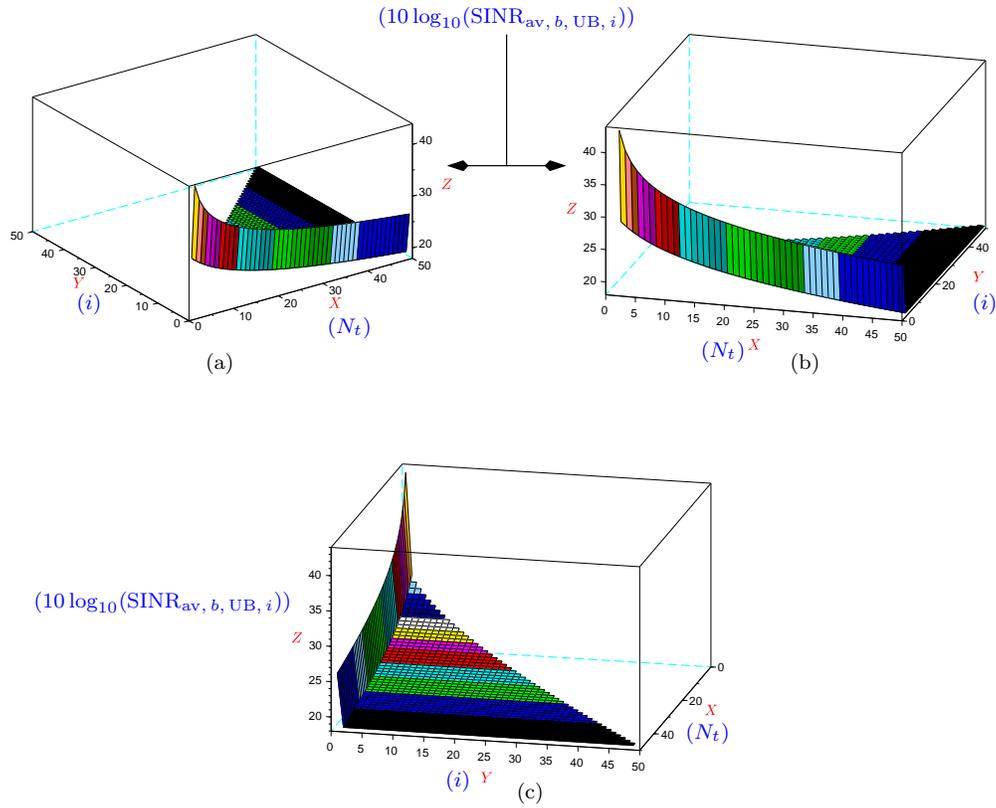}
\caption{Plot of $\mathrm{SINR}_{\mathrm{av},\, b,\,\mathrm{UB},\, i}$
         for $N_{\mathrm{tot}}=1024$, $N_{rt}=2$ after precoding. (a) Back view.
         (b) Sideview. (c) Front view.}
\label{Fig:SINR_Pre_UB_BA_Numant1024}
\end{figure*}

Similar to (\ref{Eq:SU_MMIMO_SCTC_Eq3}), consider the modified received
signal given by
\begin{equation}
\label{Eq:SU_MMIMO_PCTC_Pre_Eq1}
\tilde{\mathbf{R}}_k = \tilde{\mathbf{H}}_k
                       \tilde{\mathbf{B}}
                       \mathbf{S} +
                       \tilde{\mathbf{W}}_k
\end{equation}
where
\begin{align}
\label{Eq:SU_MMIMO_PCTC_Pre_Eq2}
\tilde{\mathbf{B}}
& = \left[
    \begin{array}{cccc}
     1                         &  0     & \cdots                 &  0\\
    \tilde{a}_{1,\, 1}         &  1     & \cdots                 &  0\\
    \vdots                     & \cdots & \cdots                 & \vdots\\
    \tilde{a}_{N_t-1,\, N_t-1} & \cdots & \tilde{a}_{N_t-1,\, 1} & 1
    \end{array}
    \right]^T       \nonumber  \\
&   \stackrel{\Delta}{=}
    \tilde{\mathbf{A}}^T
\end{align}
where $(\cdot)^T$ denotes transpose.
In (\ref{Eq:SU_MMIMO_PCTC_Pre_Eq2}), $\tilde{\mathbf{A}}$ is an $N_t\times N_t$
lower triangular
matrix with diagonal elements equal to unity and $\tilde{a}_{i,\, j}$ denotes
the $j^{th}$ coefficient of the optimum $i^{th}$-order forward prediction filter
\cite{Vasu_Book10} and $\tilde{\mathbf{B}}$ is the precoding matrix. Let
\begin{align}
\label{Eq:SU_MMIMO_PCTC_Pre_Eq3}
\tilde{\mathbf{Y}}_k & = \tilde{\mathbf{B}}^H
                         \tilde{\mathbf{H}}_k^H
                         \tilde{\mathbf{R}}_k          \nonumber  \\
                     & = \tilde{\mathbf{B}}^H
                         \tilde{\mathbf{H}}_k^H
                         \tilde{\mathbf{H}}_k
                         \tilde{\mathbf{B}}
                         \mathbf{S} +
                         \tilde{\mathbf{B}}^H
                         \tilde{\mathbf{H}}_k^H
                         \tilde{\mathbf{W}}_k.
\end{align}
Define
\begin{align}
\label{Eq:SU_MMIMO_PCTC_Pre_Eq4}
\tilde{\mathbf{Z}}_k
& = \tilde{\mathbf{H}}_k
    \tilde{\mathbf{B}}               \nonumber  \\
& = \left[
    \begin{array}{ccc}
    \tilde{Z}_{k,\,1,\, 1}    & \cdots & \tilde{Z}_{k,\, 1,\, N_t} \\
    \vdots                    & \cdots & \vdots \\
    \tilde{Z}_{k,\, N_r,\, 1} & \cdots & \tilde{Z}_{k,\, N_r,\, N_t}
    \end{array}
    \right].
\end{align}
Now \cite{Vasu_Book10}
\begin{align}
\label{Eq:SU_MMIMO_PCTC_Pre_Eq5}
\frac{1}{2}
 E
\left[
\tilde{\mathbf{Z}}_k^H
\tilde{\mathbf{Z}}_k
\right]
& = N_r
\left[
\begin{array}{cccc}
\sigma_{Z,\, 1}^2  &  0                & \cdots &  0\\
 0                 & \sigma_{Z,\, 2}^2 & \cdots &  0\\
\vdots             & \cdots            & \cdots & \vdots \\
 0                 & \cdots            &  0     & \sigma_{Z,\, N_t}^2
\end{array}
\right]                 \nonumber  \\
& \stackrel{\Delta}{=}
\tilde{\mathbf{R}}_{\tilde{\mathbf{Z}}\tilde{\mathbf{Z}}}
\end{align}
is an $N_t\times N_t$ diagonal matrix and $\sigma_{Z,\, i}^2$ denotes the
variance per dimension of the optimum $(i-1)^{th}$-order forward prediction
filter. Note that \cite{Vasu_Book10}
\begin{align}
\label{Eq:SU_MMIMO_PCTC_Pre_Eq6}
\sigma_{Z,\, 1}^2 & = \sigma^2_H                   \nonumber  \\
\sigma_{Z,\, i}^2 & \ge \sigma_{Z,\, j}^2 \qquad \mbox{for $i< j$.}
\end{align}
Let
\begin{align}
\label{Eq:SU_MMIMO_PCTC_Pre_Eq7}
\tilde{\mathbf{V}}_k & = \tilde{\mathbf{Z}}_k^H
                         \tilde{\mathbf{W}}_k             \nonumber  \\
                     & = \left[
                         \begin{array}{ccc}
                         \tilde{V}_{k,\, 1} & \cdots & \tilde{V}_{k,\, N_t}
                         \end{array}
                         \right]^T
\end{align}
which is an $N_t\times 1$ vector. Now
\begin{align}
\label{Eq:SU_MMIMO_PCTC_Pre_Eq8}
 E
\left[
\tilde{V}_{k,\, i}
\tilde{V}_{k,\, m}^*
\right] & =  E
            \left[
            \sum_{j=1}^{N_r}
            \tilde{Z}_{k,\, j,\, i}^*
            \tilde{W}_{k,\, j}
            \sum_{l=1}^{N_r}
            \tilde{Z}_{k,\, l,\, m}
            \tilde{W}_{k,\, l}^*
            \right]                             \nonumber  \\
        & = \sum_{j=1}^{N_r}
            \sum_{l=1}^{N_r}
             E
            \left[
            \tilde{Z}_{k,\, l,\, m}
            \tilde{Z}_{k,\, j,\, i}^*
            \right]
             E
            \left[
            \tilde{W}_{k,\, j}
            \tilde{W}_{k,\, l}^*
            \right]                             \nonumber  \\
        & = \sum_{j=1}^{N_r}
            \sum_{l=1}^{N_r}
             2
            \sigma^2_{Z,\, i}
            \delta_K(i-m)
            \delta_K(j-l)                       \nonumber  \\
        &   \qquad \times
             2
            \sigma^2_W
            \delta_K(j-l)                       \nonumber  \\
        & =  4 N_r
            \sigma^2_{Z,\, i}
            \sigma^2_W
            \delta_K(i-m)
\end{align}
where we have used (\ref{Eq:SU_MMIMO_PCTC_Pre_Eq5}). Let
\begin{equation}
\label{Eq:SU_MMIMO_PCTC_Pre_Eq9}
\tilde{\mathbf{F}}_k = \tilde{\mathbf{Z}}_k^H
                       \tilde{\mathbf{Z}}_k
\end{equation}
which is an $N_t\times N_t$ matrix.
Substituting (\ref{Eq:SU_MMIMO_PCTC_Pre_Eq9}) in
(\ref{Eq:SU_MMIMO_PCTC_Pre_Eq3}) we get
\begin{equation}
\label{Eq:SU_MMIMO_PCTC_Pre_Eq10}
\tilde{\mathbf{Y}}_k = \tilde{\mathbf{F}}_k
                       \mathbf{S} +
                       \tilde{\mathbf{V}}_k.
\end{equation}
Similar to (\ref{Eq:EE677A_Asgn1_Sol_Eq1}), the $i^{th}$ element of
$\tilde{\mathbf{Y}}_k$ in (\ref{Eq:SU_MMIMO_PCTC_Pre_Eq10}) is given by
\begin{equation}
\label{Eq:SU_MMIMO_PCTC_Pre_Eq11}
\tilde{Y}_{k,\, i} = \tilde{F}_{k,\, i,\, i} S_i +
                     \tilde{I}_{k,\, i} +
                     \tilde{V}_{k,\, i}
                     \quad \mbox{for $1\leq i\leq N_t$}
\end{equation}
where
\begin{align}
\label{Eq:SU_MMIMO_PCTC_Pre_Eq12}
\tilde{V}_{k,\, i}      & = \sum_{j=1}^{N_r}
                            \tilde{Z}_{k,\, j,\, i}^*
                            \tilde W_{k,\, j}                 \nonumber  \\ 
\tilde{I}_{k,\, i}      & = \sum_{\substack{j=1\\j\neq i}}^{N_t}
                            \tilde{F}_{k,\, i,\, j} S_j       \nonumber  \\
\tilde{F}_{k,\, i,\, j} & = \sum_{l=1}^{N_r}
                            \tilde{Z}_{k,\, l,\, i}^*
                            \tilde{Z}_{k,\, l,\, j}.
\end{align}
Note that from (\ref{Eq:SU_MMIMO_PCTC_Pre_Eq5}) and
(\ref{Eq:SU_MMIMO_PCTC_Pre_Eq9}) we have
\begin{equation}
\label{Eq:SU_MMIMO_PCTC_Pre_Eq13}
 E
\left[
\tilde{F}_{k,\, i,\, i}
\right] = 2 N_r\sigma^2_{Z,\, i}.
\end{equation}
Now
\begin{align}
\label{Eq:SU_MMIMO_PCTC_Pre_Eq14}
 E
\left[
\tilde{F}_{k,\, i,\, i}^2
\right] & =  E
            \left[
            \sum_{l=1}^{N_r}
            \left|
            \tilde{Z}_{k,\, l,\, i}
            \right|^2
            \sum_{m=1}^{N_r}
            \left|
            \tilde{Z}_{k,\, m,\, i}
            \right|^2
            \right]                                    \nonumber  \\
        & = \sum_{l=1}^{N_r}
            \left|
            \tilde{Z}_{k,\, l,\, i}
            \right|^4                                  \nonumber  \\
        &   \qquad +
            \sum_{\substack{m=1\\m\ne l}}^{N_r}
             E
            \left[
            \left|
            \tilde{Z}_{k,\, l,\, i}
            \right|^2
            \right]
             E
            \left[
            \left|
            \tilde{Z}_{k,\, m,\, i}
            \right|^2
            \right]                                    \nonumber  \\
        & =  4N_r(N_r+1)
            \sigma_{Z,\, i}^4.
\end{align}
Similarly
\begin{align}
\label{Eq:SU_MMIMO_PCTC_Pre_Eq15}
 E
\left[
\left|
\tilde{I}_{k,\, i}
\right|^2
\right] & =  E
            \left[
            \sum_{\substack{j=1\\j\ne i}}^{N_t}
            \tilde{F}_{k,\, i,\, j} S_j
            \sum_{\substack{l=1\\l\ne i}}^{N_t}
            \tilde{F}_{k,\, i,\, l}^* S_l^*
            \right]                                    \nonumber  \\
        & =  P_{\mathrm{av}}
            \sum_{\substack{j=1\\j\ne i}}^{N_t}
             E
            \left[
            \left|
            \tilde{F}_{k,\, i,\, j}
            \right|^2
            \right].
\end{align}
Now
\begin{align}
\label{Eq:SU_MMIMO_PCTC_Pre_Eq16}
 E
\left[
\left|
\tilde{F}_{k,\, i,\, j}
\right|^2
\right] & =  E
            \left[
            \sum_{l=1}^{N_r}
            \tilde{Z}_{k,\, l,\, i}^*
            \tilde{Z}_{k,\, l,\, j}
            \sum_{m=1}^{N_r}
            \tilde{Z}_{k,\, m,\, i}
            \tilde{Z}_{k,\, m,\, j}^*
            \right]                                    \nonumber  \\
        & = \sum_{l=1}^{N_r}
            \sum_{m=1}^{N_r}
             4
            \sigma_{Z,\, i}^2
            \sigma_{Z,\, j}^2
            \delta_K(l-m)                              \nonumber  \\
        & =  4N_r
            \sigma_{Z,\, i}^2
            \sigma_{Z,\, j}^2
\end{align}
where we have used (\ref{Eq:SU_MMIMO_PCTC_Pre_Eq5}). Substituting
(\ref{Eq:SU_MMIMO_PCTC_Pre_Eq16}) in (\ref{Eq:SU_MMIMO_PCTC_Pre_Eq15}) we get
\begin{equation}
\label{Eq:SU_MMIMO_PCTC_Pre_Eq17}
 E
\left[
\left|
\tilde{I}_{k,\, i}
\right|^2
\right]   =  4P_{\mathrm{av}} N_r
            \sigma_{Z,\, i}^2
            \sum_{\substack{j=1\\j\ne i}}^{N_t}
            \sigma_{Z,\, j}^2.
\end{equation}
Note that
\begin{equation}
\label{Eq:SU_MMIMO_PCTC_Pre_Eq18}
 E
\left[
\left|
\tilde{I}_{k,\, i} +
\tilde{V}_{k,\, i}
\right|^2
\right] =
 E
\left[
\left|
\tilde{I}_{k,\, i}
\right|^2
\right] +
 E
\left[
\left|
\tilde{V}_{k,\, i}
\right|^2
\right].
\end{equation}
The average SINR per bit for the $i^{th}$ transmit antenna is given by
(\ref{Eq:EE677A_Asgn1_Sol_Eq14}) and is equal to
\begin{align}
\label{Eq:SU_MMIMO_PCTC_Pre_Eq19}
\mathrm{SINR}_{\mathrm{av},\, b,\, i}
        & = \frac{
             E
            \left[
            \left|
            \tilde{F}_{k,\, i,\, i} S_i
            \right|^2
            \right]
            \times 2N_{rt}}
            {
             E
            \left[
            \left|
            \tilde{I}_{k,\, i} + \tilde{V}_{k,\, i}
            \right|^2
            \right]
            }                                              \nonumber  \\
        & = \frac{P_{\mathrm{av}}(N_r+1)\sigma_{Z,\, i}^2\times 2N_{rt}}
                 {P_{\mathrm{av}}
                 \sum_{\substack{j=1\\j\ne i}}^{N_t}
                 \sigma_{Z,\, j}^2+\sigma_W^2}
\end{align}
where we have used (\ref{Eq:SU_MMIMO_PCTC_Pre_Eq8}),
(\ref{Eq:SU_MMIMO_PCTC_Pre_Eq14}) and
(\ref{Eq:SU_MMIMO_PCTC_Pre_Eq17}). The upper bound on the average
SINR per bit for the $i^{th}$ transmit antenna is obtained by setting
$\sigma_W^2=0$ in (\ref{Eq:SU_MMIMO_PCTC_Pre_Eq19}) and is equal to
\begin{equation}
\label{Eq:SU_MMIMO_PCTC_Pre_Eq20}
\mathrm{SINR}_{\mathrm{av},\, b,\, \mathrm{UB},\, i}
 = \frac{(N_r+1)\sigma_{Z,\, i}^2\times 2N_{rt}}
        {
        \sum_{\substack{j=1\\j\ne i}}^{N_t}
        \sigma_{Z,\, j}^2}
\end{equation}
which is illustrated in Figure~\ref{Fig:SINR_Pre_UB_BA_Numant1024} for
$N_{\mathrm{tot}}=1024$ and $N_{rt}=2$.
The value of the upper bound on the average SINR per bit for $N_t=i=50$
is 18.6 dB. The channel correlation is given by (\ref{Eq:SU_MMIMO_SCTC_Eq46}).
\begin{figure*}[tbhp]
\centering
\input{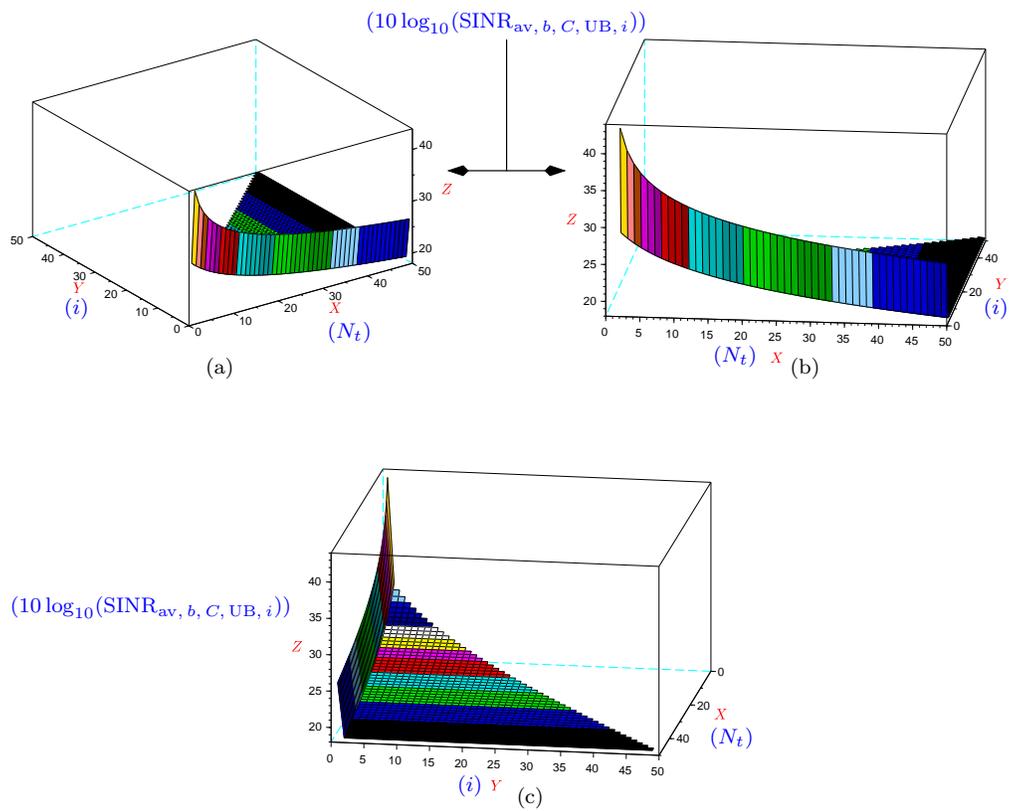}
\caption{Plot of $\mathrm{SINR}_{\mathrm{av},\, b,\, C,\,\mathrm{UB},\, i}$
         for $N_{\mathrm{tot}}=1024$, $N_{rt}=2$ after precoding. (a) Back view.
         (b) Side view. (c) Front view.}
\label{Fig:SINR_Pre_UB_AA_Numant1024}
\end{figure*}
Note that a first-order prediction filter completely decorrelates the channel
with \cite{Vasu_Book10}
\begin{align}
\label{Eq:SU_MMIMO_PCTC_Pre_Eq21}
\tilde{a}_{i,\, 1} & = -0.9 \qquad \mbox{for $1\le i\le N_t-1$}  \nonumber  \\
\tilde{a}_{i,\, j} & =  0   \qquad \mbox{for $2\le i\le N_t-1$, $2\le j\le i$}.
\end{align}
We also have \cite{Vasu_Book10}
\begin{align}
\label{Eq:SU_MMIMO_PCTC_Pre_Eq22}
\sigma_{Z,\, i}^2 & = \sigma_{Z,\, 2}^2     \nonumber  \\
                  & = \left(
                       1-|-0.9|^2
                      \right)
                      \sigma_{Z,\, 1}^2     \nonumber  \\
                  & =  0.19
                      \sigma_{Z,\, 1}^2     \qquad \mbox{for $i>2$.}
\end{align}
Therefore we see in Figure~\ref{Fig:SINR_Pre_UB_BA_Numant1024} that the
first transmit antenna $i=1$ has a high
$\mathrm{SINR}_{\mathrm{av},\, b,\,\mathrm{UB},\, i}$ due to low interference
power from remaining transmit antennas, whereas for $i\ne 1$ the
$\mathrm{SINR}_{\mathrm{av},\, b,\,\mathrm{UB},\, i}$ is low due to high
interference power from the first transmit antenna ($i=1$). The received signal
after ``combining'' is given by (\ref{Eq:SU_MMIMO_SCTC_Eq4}) and
(\ref{Eq:SU_MMIMO_SCTC_Eq34}).
Note that from (\ref{Eq:SU_MMIMO_SCTC_Eq34}) and (\ref{Eq:SU_MMIMO_PCTC_Pre_Eq12})
\begin{align}
\label{Eq:SU_MMIMO_PCTC_Pre_Eq22_1}
 E
\left[
 F_i^2
\right] & = \frac{1}{N_{rt}^2}
             E
            \left[
            \sum_{k=0}^{N_{rt}-1}
            \tilde{F}_{k,\, i,\, i}
            \sum_{l=0}^{N_{rt}-1}
            \tilde{F}_{l,\, i,\, i}
            \right]                                \nonumber  \\
        & = \frac{1}{N_{rt}^2}
            \sum_{k=0}^{N_{rt}-1}
             E
            \left[
            \left|
            \tilde{F}_{k,\, i,\, i}
            \right|^2
            \right] +
            \sum_{\substack{l=0\\l\ne k}}^{N_{rt}-1}
             E
            \left[
            \tilde{F}_{k,\, i,\, i}
            \tilde{F}_{l,\, i,\, i}
            \right]                                \nonumber  \\
        & = \frac{4N_r\sigma_{Z,\, i}^4}{N_{rt}^2}
            \sum_{k=0}^{N_{rt}-1}
            (N_r+1) +
            (N_{rt}-1) N_r                         \nonumber  \\
        & = \frac{4N_r\sigma_{Z,\, i}^4}{N_{rt}}
            (1+N_r N_{rt})
\end{align}
where we have used (\ref{Eq:SU_MMIMO_SCTC_Eq36}),
(\ref{Eq:SU_MMIMO_PCTC_Pre_Eq13}) and
(\ref{Eq:SU_MMIMO_PCTC_Pre_Eq14}). Similarly from
(\ref{Eq:SU_MMIMO_SCTC_Eq34}), (\ref{Eq:SU_MMIMO_PCTC_Pre_Eq8}),
(\ref{Eq:SU_MMIMO_PCTC_Pre_Eq17}) and (\ref{Eq:SU_MMIMO_PCTC_Pre_Eq18}) we
have
\begin{align}
\label{Eq:SU_MMIMO_PCTC_Pre_Eq22_2}
 E
\left[
\left|
\tilde{U}_i
\right|^2
\right] & = \frac{1}{N_{rt}^2}
             E
            \left[
            \sum_{k=0}^{N_{rt}-1}
            \tilde{U}_{k,\, i}'
            \sum_{l=0}^{N_{rt}-1}
            \left(
            \tilde{U}_{l,\, i}'
            \right)^*
            \right]                               \nonumber  \\
        & = \frac{1}{N_{rt}^2}
            \sum_{k=0}^{N_{rt}-1}
            \sum_{l=0}^{N_{rt}-1}
             E
            \left[
            \tilde{U}_{k,\, i}'
            \left(
            \tilde{U}_{l,\, i}'
            \right)^*
            \right]                               \nonumber  \\
        & = \frac{1}{N_{rt}^2}
            \sum_{k=0}^{N_{rt}-1}
            \sum_{l=0}^{N_{rt}-1}
             E
            \left[
            \left|
            \tilde{U}_{k,\, i}'
            \right|^2
            \right]
            \delta_K(k-l)                         \nonumber  \\
        & = \frac{1}{N_{rt}}
             E
            \left[
            \left|
            \tilde{U}_{k,\, i}'
            \right|^2
            \right]                               \nonumber  \\
        & = \frac{1}{N_{rt}}
            \left[
             E
            \left[
            \left|
            \tilde{I}_{k,\, i}
            \right|^2
            \right] +
             E
            \left[
            \left|
            \tilde{V}_{k,\, i}
            \right|^2
            \right]
            \right]                               \nonumber  \\
        & = \frac{4N_r\sigma_{Z,\, i}^2}{N_{rt}}
            \left[
             P_{\mathrm{av}}
            \sum_{\substack{j=1\\j\ne i}}^{N_t}
            \sigma_{Z,\, j}^2 +
            \sigma_W^2
            \right].
\end{align}
Substituting (\ref{Eq:SU_MMIMO_PCTC_Pre_Eq22_1}) and
(\ref{Eq:SU_MMIMO_PCTC_Pre_Eq22_2}) in (\ref{Eq:SU_MMIMO_SCTC_Eq45})
we have, after simplification, for $1\le i \le N_t$
\begin{align}
\label{Eq:SU_MMIMO_PCTC_Pre_Eq23}
\mathrm{SINR}_{\mathrm{av},\, b,\, C,\, i}
& =
\frac{2P_{\mathrm{av}}E\left[F_i^2\right]}
     {E\left[\left|\tilde{U}_i\right|^2\right]}    \nonumber  \\
& =
\frac{(N_r N_{rt}+1)\sigma_{Z,\, i}^2\times 2P_{\mathrm{av}}}
     {
      P_{\mathrm{av}}
     \sum_{\substack{j=1\\j\ne i}}^{N_t}
     \sigma_{Z,\, j}^2+\sigma_W^2}.
\end{align}
The upper bound on the average SINR per bit for the $i^{th}$ transmit antenna
is obtained by substituting (\ref{Eq:SU_MMIMO_PCTC_Pre_Eq23}) in
(\ref{Eq:SU_MMIMO_SCTC_Eq45_1}) and is equal to
\begin{align}
\label{Eq:SU_MMIMO_PCTC_Pre_Eq24}
\mathrm{SINR}_{\mathrm{av},\, b,\, C,\, \mathrm{UB},\, i}
& =  \frac{(N_r N_{rt}+1)\sigma_{Z,\, i}^2\times 2}
     {
     \sum_{\substack{j=1\\j\ne i}}^{N_t}
     \sigma_{Z,\, j}^2}                   \nonumber  \\
&    \approx
     \mathrm{SINR}_{\mathrm{av},\, b,\, \mathrm{UB},\, i}
\end{align}
for $1\le i \le N_t$, $N_r \gg 1$. This is illustrated in 
Figure~\ref{Fig:SINR_Pre_UB_AA_Numant1024} for $N_{\mathrm{tot}}=1024$ and
$N_{rt}=2$. We again observe that the first transmit antenna ($i=1$) has a high
upper bound on the average SINR per bit, after ``combining'', compared to the
remaining transmit antennas.
The value of the upper bound on the average SINR per bit  after ``combining''
for $N_t=i=50$, $N_{\mathrm{tot}}=1024$ is 18.6 dB.
After concatenation, $\tilde{Y}_i$ for $0\le i\le L_d-1$,
in (\ref{Eq:SU_MMIMO_SCTC_Eq4}) and (\ref{Eq:SU_MMIMO_SCTC_Eq34}) is
given to the turbo decoder \cite{Vasu_Book10,KV_OpSigPJ2019}.
Let (see (26) of \cite{KV_OpSigPJ2019}):
\begin{align}
\label{Eq:SU_MMIMO_PCTC_Pre_Eq25}
\tilde{\mathbf{Y}}_1 & = \left[
                         \begin{array}{ccc}
                         \tilde{Y}_0         & \ldots & \tilde{Y}_{L_{d1}-1}
                         \end{array}
                         \right]                      \nonumber  \\
\tilde{\mathbf{Y}}_2 & = \left[
                         \begin{array}{ccc}
                         \tilde{Y}_{L_{d1}}  & \ldots & \tilde{Y}_{L_{d}-1}
                         \end{array}
                         \right].
\end{align}
Then \cite{Vasu_Book10,KV_OpSigPJ2019}
\begin{align}
\label{Eq:SU_MMIMO_PCTC_Pre_Eq26}
\gamma_{1,\, i,\, m,\, n} & =
                            \exp
                            \left[
                             -
                            \,
                            \frac{
                            \left|
                            \tilde Y_i - F_i S_{m,\, n}
                            \right|^2}{2\sigma^2_{U,\, i}}
                            \right]         \nonumber  \\
\gamma_{2,\, i,\, m,\, n} & =
                            \exp
                            \left[
                             -
                            \,
                            \frac{
                            \left|
                            \tilde Y_{i1} - F_{i1} S_{m,\, n}
                            \right|^2}{2\sigma^2_{U,\, i}}
                            \right]
\end{align}
where
\begin{equation}
\label{Eq:SU_MMIMO_PCTC_Pre_Eq27}
i1 = i + L_{d1}  \qquad \mbox{for $0 \le i \le L_{d1}-1$}.
\end{equation}
The rest of the turbo decoding algorithm is similar to that discussed
in \cite{Vasu_Book10,KV_OpSigPJ2019} and will not be repeated here.
In the next subsection we present the computer simulation results for correlated
channel with precoding and PCTC.
\subsection{Simulation Results}
\label{SSec:PCTC_Corr_Results}
\begin{figure*}[tbhp]
\centering
\input{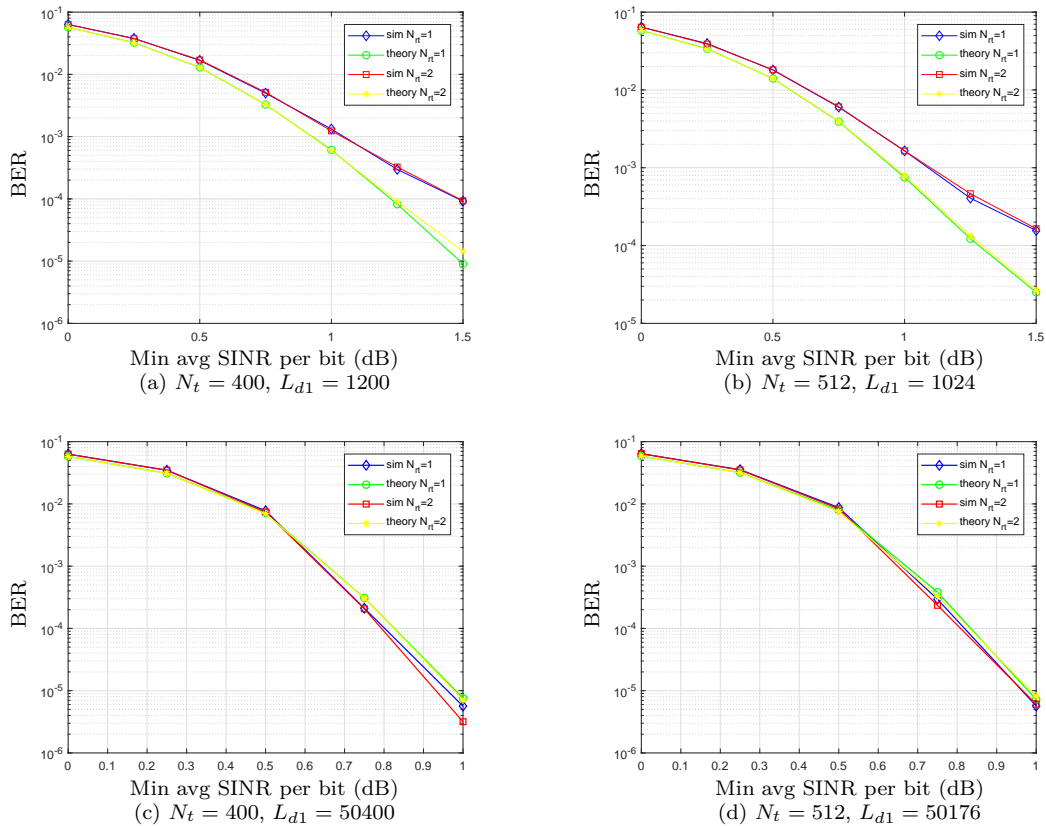}
\caption{Simulation results with precoding for $N_{\mathrm{tot}}=1024$.}
\label{Fig:Ber_NUM_ANT1024_Corr_DMT51_2_i}
\end{figure*}
The channel correlation is given by (\ref{Eq:SU_MMIMO_SCTC_Eq46}).
The BER results for $N_{\mathrm{tot}}=1024$ with precoding are depicted in
Figure~\ref{Fig:Ber_NUM_ANT1024_Corr_DMT51_2_i}. Incidentally, the value of the
upper bound on the average SINR per bit before and after ``combining'' for
$N_t=i=512$, $N_{\mathrm{tot}}=1024$ is 6 dB.
\begin{figure*}[tbhp]
\centering
\input{ber_num_ant32_corr_dmt51_2_i.pstex_t}
\caption{Simulation results with precoding for $N_{\mathrm{tot}}=32$.}
\label{Fig:Ber_NUM_ANT32_Corr_DMT51_2_i}
\end{figure*}
The BER results for $N_{\mathrm{tot}}=32$ with precoding are depicted in
Figure~\ref{Fig:Ber_NUM_ANT32_Corr_DMT51_2_i}. Note that since the average
SINR per bit depends on the transmit antenna, the \textit{minimum} average
SINR per bit is indicated along the $x$-axis of
Figures~\ref{Fig:Ber_NUM_ANT1024_Corr_DMT51_2_i} and
\ref{Fig:Ber_NUM_ANT32_Corr_DMT51_2_i}. We also observe from
Figures~\ref{Fig:Ber_NUM_ANT1024_Corr_DMT51_2_i}(a, b) and
\ref{Fig:Ber_NUM_ANT32_Corr_DMT51_2_i} that there is a large difference
between theory and simulations. This is probably because, the
average SINR per bit is not identical for all transmit antennas. In particular,
we observe from Figures~\ref{Fig:SINR_Pre_UB_BA_Numant1024} and
\ref{Fig:SINR_Pre_UB_AA_Numant1024} that the first transmit antenna
has a large average SINR per bit compared to the remaining antennas.
However, in Figures~\ref{Fig:Ber_NUM_ANT1024_Corr_DMT51_2_i}(c, d) there is
a close match between theory and simulations. This could be attributed to
having a large number of blocks in a frame, as given by
(\ref{Eq:SU_MMIMO_SCTC_Eq2}), resulting in better statistical properties. 
Even though the number of blocks is large in
\ref{Fig:Ber_NUM_ANT32_Corr_DMT51_2_i}, the number of transmit antennas is small,
resulting in inferior statistical properties. In order to improve the accuracy
of the BER estimate for $N_{\mathrm{tot}}=32$, we propose to transmit
``dummy data'' from the first transmit antenna and ``actual data'' from the
remaining antennas. The BER results shown in
Figure~\ref{Fig:Ber_NUM_ANT32_Corr_DMT51_3_i} indicates a good match between
theory and practice. However, comparison of
Figures~\ref{Fig:Ber_NUM_ANT1024_Corr_DMT51_2_i} and
\ref{Fig:Ber_NUM_ANT1024_Corr_DMT51_3_i} demonstrates that ``dummy data'' is
ineffective for large number of transmit antennas.
\begin{figure*}[tbhp]
\centering
\input{ber_num_ant1024_corr_dmt51_3_i.pstex_t}
\caption{Simulation results with precoding and dummy data for
         $N_{\mathrm{tot}}=1024$.}
\label{Fig:Ber_NUM_ANT1024_Corr_DMT51_3_i}
\end{figure*}
\begin{figure*}[tbhp]
\centering
\input{ber_num_ant32_corr_dmt51_3_i.pstex_t}
\caption{Simulation results with precoding and dummy data
         for $N_{\mathrm{tot}}=32$.}
\label{Fig:Ber_NUM_ANT32_Corr_DMT51_3_i}
\end{figure*}

\section{Conclusions \& Future Work}
\label{Sec:Conclude}
This article presents the advantages of single-user massive multiple
input multiple output (SU-MMIMO) over multi-user (MU) MMIMO systems. The
bit-error-rate (BER) performance of SU-MMIMO using serially concatenated
turbo codes (SCTC) over uncorrelated channel is presented. A semi-analytic
approach to estimating the BER of a turbo code is derived. A detailed
signal-to-interference-plus-noise ratio analysis for SU-MMIMO over
correlated channel is presented. The BER performance of SU-MMIMO with
parallel concatenated turbo code (PCTC) over correlated channel is studied.
Future work could involve estimating the MMIMO channel, since the
present work assumes perfect knowledge of the channel.


%





\ifCLASSOPTIONcaptionsoff
  \newpage
\fi

\begin{footnotesize}

\bibliographystyle{IEEEtran}
\bibliography{mybib,mybib1,mybib2,mybib3,mybib4,mybib5}
\end{footnotesize}
\end{document}